\documentclass[12pt]{article}

\usepackage{graphicx,amsmath,amsfonts,amssymb,amsthm}

\theoremstyle{plain}

\newtheorem{proposition}{Proposition}

\theoremstyle{remark}
\newtheorem{remark}{Remark}
\newtheorem{example}{Example}
\newtheorem{definition}{Definition}

\begin{document}

\title{Main Physical Aspects of the Mathematical Conception of Energy in Thermodynamics}

\author{\textbf{V. P. Maslov}}

\date{\small National Research University Higher School of Economics, Moscow, 123458, Russia;\\
Department of Physics,  Lomonosov Moscow  State  University, Moscow, 119234, Russia; \\
Ishlinsky Institute for Problems in Mechanics,\\
Russian Academy of Sciences, Moscow, 119526, Russia; \\
v.p.maslov@mail.ru }

\maketitle

\begin{abstract}
We consider the main physical notions and phenomena described by the author in his mathematical
theory of thermodynamics. The new mathematical model yields the equation of state for a wide class
of classical gases consisting of non-polar molecules provided that the spinodal, the critical
isochore and the second virial coefficient are given. As an example, the spinodal, the critical
isochore and the second virial coefficient are taken from the Van-der-Waals model. For this specific
example, the isotherms constructed on the basis of the author's model are compared to the
Van-der-Waals isotherms, obtained from completely different considerations.

\end{abstract}

\section{Introduction}

In the present paper, we present the foundations of the new mathematical model of thermodynamics for
the values of the energy for which the molecules  are in the pre-plasma
state, in particular  as  pressure $P$ is  near-zero.

The new mathematical model constructed by the author in the cycle of papers
\cite{MTN_94-5}--\cite{RJMP_21-2},  differs somewhat from the commonly accepted model of
phenomenological thermodynamics and allows us to construct the equation of state for a wider class
of classical gases.

We mainly study the metastable states in the case of gases and in the case of fluids up to the
critical isochore $\rho=\rho_c$.  The liquid isotherms are divided into two parts: the region with
temperatures near the critical temperature $T_c$ and the region of the ``hard'' liquid whose
isotherms pass through the point of zero pressure.

Further, the distributions in the Van-der-Waals model are compared to the distributions appearing in
the author's new model for classical gases with given critical temperatures, in particular, with
Van-der-Waals critical temperature. The precision with which the isotherms constructed from the
author's theory coincide with the Van-der-Waals isotherms turns out to be less than the accepted
precision in experimental studies.  The probability of such a coincidence being accidental is
``infinitely small''.

All the distributions of energy levels for classical gas considered in the present paper have been
established mathematically, see~\cite{MTN_94-5}--\cite{Elsevier_2014} and references  therein) and
have met  with  approval (see, for example, the review~\cite{Teor_ver_Premia}, published in the
journal ``Probability Theory and its Applications'' in connection with the State Prize of the
Russian Federation in Science and Technology granted to the author in 2013 (for developing the
mathematical foundations of modern thermodynamics).

\section{Van-der-Waals Normalization and the Law of \\ Corresponding States}

The correct choice of units of measurement for different gases allowed Van-der-Waals to compare the
parameters of the gases to dimensionless quantities, thus obtaining the famous ``law of
corresponding states''.

For every gas, there exists a critical temperature $T_c$ and a critical pressure $P_c$ such that if
the temperature or the pressure is greater than their critical values, then gas and liquid can no
longer be distinguished. This state of matter is known as a fluid.

The Van-der-Waals ``normalization'' consists in taking the ratio of the parameters by their critical
value, so that we consider the reduced temperature $T_r=T/T_c$ and the reduced pressure$P_r=P/P_c$.
In these coordinates, the diagrams of the state of various gases resemble each other -- this a
manifestation of the law of corresponding states.

The equation of state for the Van-der-Waals gas in dimensionless variables has the form:
\begin{equation}\label{vdv}
(P_r+\alpha\rho^2)(1-\tilde{\alpha}\rho) = \rho T_r,
\end{equation}
where   $\alpha={3^3}/{2^6}$,  $\tilde{\alpha}={1}/{2^3}$,  and   the dimensionless  concentration
(density)  $\rho=N/V$  is chosen  so that  $\rho_c=8/3$ in the  critical point.

In thermodynamics, the dimension of the internal energy $E$  is equal to the dimension of the
product $PV$ of the pressure by the volume. On the other hand, temperature plays the role of mean
energy $E/N$ of the gas (or of the liquid). For that reason, when the dimension of temperature is
given in energy units,  the parameter  $Z=P/\rho T$, known as the  \emph{compressibility factor}, is
dimensionless. For the equation of state of the gas or liquid, the most natural diagram is the
$P_r$-$Z$ diagram, in which the dimensionless pressure $P_r$ is plotted along the $x$-axis and the
dimensionless parameter $Z$, along the $y$-axis. A diagram with axes $P_r, Z$  is called a
\emph{Hougen--Watson diagram}. In the Van-der-Waals model, the critical value of the compressibility
factor is
$$
Z_{c}=\frac{P_{c}V_{c}}{NT_{c}}= \frac 38.
$$

\begin{figure}[h!]
\centering
\includegraphics[draft=false]{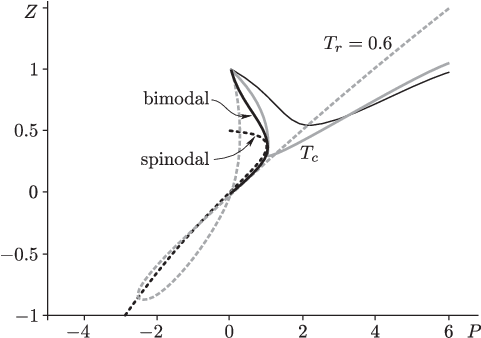}
\caption{ Van-der-Waals isotherms on the $P_r$-$Z$ diagram. On the plot   $P=P_r$, $T=T_r$. The
curve shown by the fat hashed line is the critical isotherm $T_r=1$.  The dotted line is the
spinodal, the thin hashed line is the binodal.}. \label{mas_00}
\end{figure}

\begin{figure}[h!]
\centering
\includegraphics[draft=false]{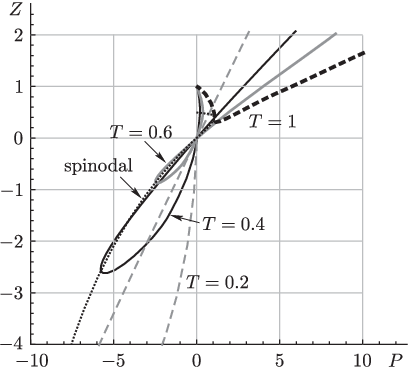}
\caption{ The $P_r$-$Z$ diagram for the Van-der-Waals equation with $P=P_r$. }
\label{mas_000}
\end{figure}

In Figs.~\ref{mas_000}--\ref{mas_00}, the Van-der-Waals isotherms are shown in a $P_r$-$Z$
Hougen--Watson diagram, for which we used the normalization $P_r=P/P_c$,  $T_r=T/T_c$,  where $T_c$
is the critical temperature.

\section{On the Number of Degrees of Freedom and the Parition Theory of Integers}

In the physics literature, the notion of number of degrees of freedom is used for ideal gases, in
which particles do not interact. In probability, the number of degrees of freedom is also considered
independently of any interaction.

When physicists speak of the number of degrees of freedom, they  usually have in mind the number of
degrees of freedom of a single molecule. Normally, a one-atom molecule will have 3 degrees of
freedom, a two-atom molecule, 5. In a gas, the  molecules move with different speeds and different
energies; although their mean energy -- temperature -- is the same, their individual energies may be
different, and the number of degrees of freedom of different molecules can also differ. Thus a
two-atom molecule with very high energy may have a number of degrees of freedom greater than than 5,
and this will affect the mean value (average over all molecules) of the number of degrees of freedom
of the given gas. This mean value is called the {\it collective number of degrees of freedom} of the
gas, and is a fractional rather than a whole number.

A one-atom molecule has 3 degrees of freedom, a two-atom molecule, 5. Two-atom molecules are
regarded in~\cite{Landau_StPhys} as molecules of ideal gas. Nevertheless, saying that the number of
degrees of freedom of a molecule is 5 means that we are implementing an exclusion rule: we are
saying that one of the degrees of freedom is excluded. More precisely, we regard the molecule as a
dumbbell, and exclude its oscillations as a rod. However, if the temperature increases, these
``rod'' oscillations will take place and so, for sufficiently high temperatures (energies), the
molecules will have a greater fractional  number of degrees of freedom.

In the author's model, the number of degrees of freedom is an important independent parameter, so
that the notion of exclusion need not be defined by means of some internal considerations
\footnote{Sometimes, instead of the number of degrees of freedom, such notions as free will or
freedom of choice and so on are considered (see, e.g. \cite{Kad_Dyn}).}.

Although the collective number of degrees of freedom of a classical gas can be determined by the
initial interaction between its particles, in the probabilistic description of the generalized ideal
gas (a notion that we present in Section 5), this parameter is introduced independently.

If nuclear forces are taken into consideration, it can be shown that a wide class of two-atom
molecules have 5 degrees of freedom, but this is difficult to establish. And so we prefer to
stipulate axiomatically, as in probability theory, that on a given energy level there can be only
one or, say, no more than $K$ particles.

Another important consideration is that we do not take into account the number of particles and the
volume separately, we consider their ratio -- density (concentration).

The consideration of density alone leads to the following important consequence in thermodynamics:
density does not depend on the numeration of particles contained in the given volume. Whatever the
numeration of the particles, the density remains the same. It is commonly believed (in particular,
it is stated in the book~\cite{Landau_Kv_Mech}) that classical particles differ from quantum
particles in that they can be numbered and then the motion of any individual particle can be
followed by keeping track of its number. This is correct, but if the behavior of a multi-particle
system is described by equations involving density, e.g. via a probability distribution, then such a
description does not depend on the numbering of particles.

Therefore, all the results from quantum mechanics that follow from the remarkable fact that the
solution does not depend on numeration can be carried over to the description of any classical
multi-particle system by means of equations containing density\footnote{In particular, the most
important thing in the chapter on identical particles in~\cite{Landau_Kv_Mech} is the method of
secondary quantization introduced by Dirac and neatly explained in mathematical terms by Fock (the
Fock space, the creation and annihilation operators and so on). Since density does not depend on
numeration,  the method of secondary quantization had been carried over to classical multi-particle
systems described by equations depending on density. This was done by Schoenberg in 1953--54. Later
Schoenberg's approach was  generalized  in the author's joint article with O.Yu.
Shvedov~\cite{Masl-Shv_Com-germ}.}.

As an illustrative example, we consider the purchase of 1 kilogram of granulated sugar.
If this amount of sugar is measured by calculated the grains
(and thus the grains must be first numerated),
then it is obvious that this process takes a lot of time.
Assume that the buyer has a fixed time for the measurement $t_{mes}$
which is equal to one hour and there is no balance to weight the sugar.
Then the buyer will measure the granulated sugar by using some vessels.
The results of weighting and of measuring by vessels
does not change if two grains in the measured volume interchange their places.
This means that (1) \emph{the particle identity principle holds for the sugar grains}
and (2) \emph{the sum of particles is independent of their location},
i.e., the arithmetic property is satisfied~\cite{MTN_102-6-shcom}.
In particular, this implies that one can use a Bose--Einstein-type distribution
in this ``classical'' situation.

The exclusions that we impose will lead us to a more general distribution in which we stipulate
that no more than $K$  particles are on the same energy level. The natural number $K$ will be called
the \emph{maximal occupation number} or  \emph{maximal admissible size of clusters}.

The corresponding energy distribution, often called parastatistical~\cite{Gentile}--\cite{RJ_21-1},
will be described in the next sections.
To specific and generalize the parastatistical distributions,
we shall use the above-listed items~(1) and~(2).

\begin{remark}
In the Landau--Lifshits textbook~\cite{Landau_StPhys}, two relations are given as the basic equations
\begin{align}\label{L-L}
\sum_j N_j=N,
\\
\sum_j \varepsilon_j N_j={M}, \nonumber
\end{align}
where ${M}$ is the energy, $N$ is the number of particles and $\{\varepsilon_j\}$ is a
discrete family of energy levels (see \cite{Vershik},~\cite{MTN_Mas_Naz_99-1}).

The values $\varepsilon_j$  in Eq.~\eqref{L-L} are related to the existing interaction
potential for the particles. It turns out that if we use the interaction with collective number of
degrees of freedom equal to $D$, then we must put\footnote{The Landau--Lifshits book gives
a more general system of equations, later specified in~\cite{Auluck}, \cite{Agarwala}:
$\sum_j \,\varepsilon_jN_j=E,$
$\varepsilon_j=j^{2/D}$.
For Landau $M=E$ is energy.}
\begin{equation}\label{ep-1}
\varepsilon_j= \text{const} \cdot  j^{D/2},
\end{equation}
i.e.,
\begin{equation}\label{ep-2}
{M}=\sum_j \text{const} \cdot j^{D/2} N_j.
\end{equation}
This statement is a theorem proved in the author's paper~\cite{MTN_98-1}.
\end{remark}

 The first person who noticed the unexpected relationship between
the partition problem in number theory (in particular the famous
discovery of Ramanujan) and nuclear disintegration, was Niels Bohr,
the founder of quantum theory (see~\cite{MTN_102-2}).

 It is well known that Bohr's most famous opponent was the great
Albert Einstein himself.
 The polemic between Bohr and Einstein,
during which Bohr gave convincing answers to all of Einstein's
counterarguments, only fortified Bohr's position and helped him
in developing his quantum approach.
 In a commentary to one of
Einstein's letters, Max Born wrote: ``...
 Einstein's attitude to
quantum mechanics was a heavy blow to me, he refuted the theory
without any argumentation, only referring to his `inner voice'...''
\cite{Ein_Born}.
 It was only after the experimental corroboration of the
disintegration of heavy Lithium~\cite{Litii} and Uranium-235 that Einstein
did support, albeit reluctantly, the A-bomb project.

The Bohr--Kalckar paper~\cite{Bohr_Kal} of 1937 was the first work in which the coincidence of
``thermodynamical formulas'' with the number-theoretic formulas for partitions of integers into sums
was discovered. Here is a quotation from that paper. ``Under the simplifying assumption that each
level is the combination of a certain number of quantities assuming nearly equidistant values, one
can easily calculate the density of nuclear levels under high perturbations. Denote by $p(M)$ the
number of possible ways of presenting a positive integer as a sum of smaller positive integers. An
asymptotic formula for $p(M)$ was obtained by G.~Hardy and S.~Ramanujan. For large values of $M$
this formula may be approximately written in the form
$$
p(M)\sim \frac{1}{4\sqrt{3}M} e^{\pi \sqrt{\frac{2}{3}M}}.
$$
 Let us choose
$2\cdot 10^5 e V$
for the unit of energy, which
approximately corresponds to the common distance between the lowest
levels of the heavier nuclei.
 For the number of partitions for
which the perturbation energy of
$8\cdot 10^6 e V$
will be
obtained, we will then find
$p(M)\sim 2\cdot 10^4$.
 This means that
the mean distance between levels approximately equals
$10eV$,
which
roughly corresponds to the densities of distribution of levels
calculated from the collisions of slow neutrons.
$\langle\dots\rangle
$ The formulas for the density of nuclear
levels obtained by analogy with thermodynamics practically
coincide, at least in the exponential dependence on the total
energy of excited nuclei, with the formula for
$p(M)$
if one
understands the number
$M$
as the measure of total energy expressed
as the difference of energies between the lowest levels taken for
the unit of measure'' (see pp. 337--338 in the Russian translation~\cite{Bohr_Kal}).

 Bohr's and Kalckar's considerations about the distance between lowest
energy levels is correct, because for the Schr\"{o}dinger equation,
the equations of the lower levels in the potential trough are
quadratic.

 Note that the probability of the coincidences indicated by Bohr
and Kalckar being accidental is practically zero.

 Later Bohr proposed using Uranium-235 as an example of the nucleus
of an atom, since it was most appropriate for his construction and
used it to explain how to generate a nuclear reaction~\cite{Bohr_Wheeler}.

 The historian and archivist of the British Committee for Atomic
Energy Margaret Going, on the basis of her study of the opinions of
authoritative physicists, wrote: ``...
 The work of English and
American scientists on the A-bomb came directly from articles of Professor Bohr and Doctor
Wheeler...''. Thus we may conclude that the creation of the A-bomb was based on the fantastic
discovery of Ramanujan.

 An essential role in establishing the new world outlook in physics
was played by the numerous volumes of the famous treatise by Landau
and Lifshits, in particular,~\cite{Landau_StPhys} and \cite{Lif_Pit-2}.
 In the book~\cite{Landau_StPhys} in
Section~40 ``Nonequilibrium ideal gas,'' the authors present the  system of equations~\eqref{L-L}  and \eqref{ep-1}--\eqref{ep-2}, on
which their further exposition is based.
 Both
$N$
and
$N_j$
in these equations are integers.
 And so
equations~\eqref{L-L} coincide with the Diophantine equations for
partitions in number theory.
 In the subsequent sections of the book~\cite{Landau_StPhys}
 the main equations of thermodynamics are obtained {\it without
appealing to the so-called three main principles of
thermodynamics}, which appear in all textbooks on thermodynamics,
but not in~\cite{Landau_StPhys}.

 Bohr's paper~\cite{Bohr_Kal}  generated only a trickle of mathematical and
physical papers, in which the connection between the statistics of
Bose--Einstein and Fermi--Dirac and the Ramanujan formula was
studied in more detail~\cite{Auluck}--\cite{Rovenchak}.
 Thus papers on this topic appeared
in the mathematical journal ``Mathematical Proceedings of the Cambridge Philosophical Society.'' It
was shown that if one takes into account the connections with partition theory, the Ramanujan
formula and the Hardy--Ramanujan theorem in number theory, then it becomes clear that, in the case
when there are repeated summands in the partition, the leading term of the Ramanujan formula
coincides with the entropy of the Bose--Einstein statistics, whereas in the case when there are no
repeated summands, this term coincides with the Fermi--Dirac entropy (bosons and fermions~-- no
other particles are observed).
 Apparently, one of the last papers in this series is~\cite{Rovenchak}.
 It
contains a bibliography of the topic.
 But overall it seems that
contemporary physicists did not pay attention to that series of
papers and forgot that they are indebted to Ramanujan for the
remarkable revolution in the scientific world outlook.

 It follows from Bohr's paper that his model of the nucleus of an
atom does not involve the interaction of particles in the form of
attraction.
 When Bohr visited the USSR in 1961, he demonstrated a
simple model of a nucleus consisting of little balls in a cup.
 Several balls were placed in a cup, and then another little ball,
endowed with a certain energy, was slipped into the cup.
 Were the
cup empty, the new ball would have slipped out of the cup, but it
shared some of its energy with the other balls and stayed in the
cup.
 This is due to the fact that all the balls were in a common
potential field.
 Thus Bohr regarded the nucleus as consisting of
particles that do not attract each other.

 Bohr's model of the nucleus is a model without attraction of
molecules, and so is Frenkel's model of liquid (liquid drop model).
 Such a model of the nucleus appears to contradict the physical viewpoint as well
 as the  commonsense (or naive) point of view.
 Nevertheless, it adequately
describes nuclear fission, and Frenkel's model adequately describes
the behavior of liquids.
 The Frenkel model contains ``holes'' but
does not involve mutual attraction of molecules.

 Incidentally, Frenkel subtly complained about the difficulties of
overcoming traditional viewpoints, using the word ``we,'' i.e.
``physicists.''   He writes: ``We easily get used to uniform and steady
things, we stop noticing them.
 What is habitual seems
understandable, but we don't understand the new and unusual, it
seems unnatural and obscure
$\langle\dots\rangle$.
 Essentially, we
never really understand, we can only get used to''~\cite[p.~63]{Frenkel}.
 If we follow Bohr, we see that in the thermodynamics that he talks
about (see above), there is no attraction between molecules, but
there is a common potential, in particular the Earth's
gravitational attraction.
 Molecules can collide, as they do in
ideal gas, but there is no mutual attraction.

 In Bohr's paper mentioned above, he noted the connection between
nuclear fission and the theory of partitions in number theory.
 On
the other hand, he proposed the drop liquid model of the
nucleus.
 This leads us to the idea that the theory liquids is also
connected to partition theory.
 In that connection Bohr writes only
about the {\it analogy} with thermodynamics, and so, like his pupil
Landau, this means that they did not have in mind the old
thermodynamics based on the ``three fundamental principles,'' but
the thermodynamics understood in the framework of the new world
outlook based on Bohr's quantum theory postulates and partition
theory.

 The author, following the way traced out by Bohr and Landau, has
developed a mathematical approach based on number theory problems
and quantum theory.
 In the author's model there is no attraction
between particles\footnote{To be more precise, attraction between
particles is used only at the boundary conditions in fixing the
experimental values of the first and second virial coefficient of
the given gas.
 Also note that thermodynamics is constructed with
``logarithmic accuracy.''},
but there is a common potential, in particular, the potential due
to the Earth's gravitation and its rotation.

 The author has shown that his approach to thermodynamics yields the
same formulas as partition theory with logarithmic precision (which
means the formulas coincide from the point of view of tropical
arithmetic).
 The notion of logarithmic accuracy was introduced in
Vol.~X of the Landau--Lifshits textbooks~\cite{Lif_Pit-2} on p.~211.
 It is
defined as follows: logarithmic accuracy of
$M$
means that we find
$M$
not up to
$o(M)$,
but only up to
$o($ln$(M)$.

 Before passing to the formulation of the main formulas and results,
we recall the notions of tropical arithmetic and logarithmic
accuracy.

 The arithmetics of a system may be decimal or binary.
 The well-known notion of integer part of a real number,
\textit{entier}, is denoted by square brackets:
$[a]$~stands for
the maximum integer that does not exceed a given real number~$a$.
 It is this integer that is usually retained in human memory.
 For
example, although the number~$2.99$
is close to~$3$,
our memory
retains its integer part~$2$;
this peculiarity is used by marketing
people when assigning prices to commodities in shops.

 From the viewpoint of arithmetics, it is quite natural to discard
the fractional part for sufficiently large~$a$.
 If the numbers are
as large as is common for macroscopic systems, it is convenient to
use a generalization of this notion for large numbers.
 By
$[a]_{10}$
we denote decimal arithmetics, where
$[\cdot]_{10}$
means the same entier for rationals but with respect to~$10$.
 For
example,
$[15]_{10}=10$,
$[90]_{10}=10$,
$[105]_{10}=100$,
and
$[6\cdot 10^{23}]_{10}=10^{23}$.

 For the sum, we have
$[10^{15}+10^{14}]_{10}=10^{15}$;
i.e., the
sum equals
$\max(10^{15}, 10^{14})$.
 For the product,
$10^{15}
\cdot 10^{14}= 10^{29}$;
i.e., the product is equal to the number
of zeros in a multidigit number.

 Similar rules hold for a system considered in the binary
arithmetics
$[a]_{2}$.

 The arithmetics thus constructed not only corresponds to the
$(\max, +)$-algebra but also takes into account precision and
neglects
$c$-numbers, just as it happens in the Maslov--Litvinov dequantization.
 Hence we can say that this arithmetics
is ``tropical.''

 For the natural logarithm, we deal with a special situation.
 Here
we must use the great formula due to the great Ramanujan and the
natural logarithm of the solution given by this formula.
 By
considering the integer parts
$[\log_e p(M)]$,
we obtain a
partition of
$p(M)$
into similar subsets of integers.

 Note that the use of the natural logarithm is natural if the
formulas use the polylogarithm, because the polylogarithm emerges
from the Stirling formula, and the Stirling formula contains
Euler's number~$e$.

 Thus, in addition to the arithmetics described by the entiers
$[\cdot]_{10}$
and
$[\cdot]_{2}$,
we introduce the arithmetics
$[\cdot]_{e}$
on the basis of partition theory.

\begin{definition}
 We say that two numbers coincide with logarithmic accuracy if the
values of their normal arithmetics
$[\cdot]_{e}$
coincide.
 In
other words,
$\ln A\cong \ln B$
means that
$\ln A=\ln B+o(\ln B)$.
\end{definition}

 Since this arithmetics puts all elements coinciding with each other
with ``logarithmic accuracy'' into the same equivalence class, we
see that convergent series are equivalent to polynomials.
 It
follows that, to each series
$\Phi$,
we can assign the degree of
the corresponding polynomial.
 We denote this number by
$n(\Phi)$.
 If one can assign an enveloping series to a divergent series, then
the minimum number of the enveloping series naturally determines
the degree
$n(\Phi)$
of the corresponding polynomial.

Now we  can briefly present a classical problem in
partition theory of integers renewed by the inclusion of the number
zero and negative entropy (negentropy).

First, consider some examples.

\begin{example}
 If we partition the number 5 into 2 summands, zero summands not
allowed, then we obtain
$p_N(M) =2$,
5=4+1=3+2, where
$N$
is the
number of summands,
$N= 1, 2, \dots, M$,
and
$p_N(M)$
is the number
of partitions of~$M$
into~$N$
summands.
 There are two possible
partitions in this example for
$N=2$.
\end{example}

\begin{example}
 Let us allow zero summands in partitions of~$M$
into~$N$
summands.
 Then
$p_N(5) =3$
for
$N=2$
in the preceding example,
5=5+0=4+1=3+2.
 All in all, there are three possible partitions in
this case.

 When partitioning 5 into 3 with zeros allowed, the
$p_N(M)$
partitions without zeros,
$ 5=3+1+1= 2+2+1= \dots$ , are supplemented
with partitions containing zeros, $4+1+0= 3+2+0=5+0+0$.
 With zeros
taken into account, all preceding partitions are repeated; i.e., we
obtain the sum of all partitions without zeros.
\end{example}

 We introduce the following notation:

$\bar{p}_s(M)=\sum_{i=1}^s p_i(M)$.

\begin{proposition}
 The number of partitions of~$M$
into~$N$
summands, zero summands
allowed, coincides with
$\bar{p}_s(M)$
for
$s=N$.
\end{proposition}

\begin{definition}\label{def-1}
 Let
$q_N(M)$
be the number of partitions of~$M$
into~$N$
distinct
summands, and let
$$
q(M)=\sum_N q_N(M).
$$
\end{definition}

 By definition, the partition 3+1+1 of~5 into~3 terms is excluded
from~$q_N(M)$.
 All the zeros are excluded as well.

 Let us continue the partition~$\bar{p}_N(M)$
by the
numbers~$q_N(M)$.
 This continuation is continuous, because for
$M=1$
there are two partitions,
$M=1=1+0$,
taken into account in
$\bar{p}_N(M)$
and one partition,
$q_N(1)=1$,
taken into account in
$q_N(M)$.

 We define \textit{natural entropy} as the natural logarithm of
$\bar{p}_N(M)$
and \textit{natural negentropy} as the natural
logarithm of
$q_N(M)$.
 The passage through the point
$M=1$
is the
passage from positive natural entropy into negative natural
negentropy
$\log_e(q_N(M))$.

 Since
$p_1(M)=1$
and
$p_M(M)=1$,
it follows that there exist maxima
between
$N=1$
and
$N=M$.
 Of these maxima, take the maximum value~$\bar{N}$.

 Since
$p_N(M)$
decreases for
$N>\bar{N}$
and
$\bar{p}_N(M)$
is
nondecreasing, we see that the increase in
$\bar{p}_N(M)$
can only
be due to the increase in the number of zeros.

\section{The Equation of State}

In the simplest version of thermodynamics, one considers the conjugate extensive-intensive pairs
pressure-volume, temperature-entropy, chemical potential-number of particles.

One of the main notions of thermodynamics is the equation of state. In 6-dimensional phase space,
where the intensive thermodynamical variables $P$, $T$, and $-\mu$ play the role of coordinates and
the corresponding extensive variables $V$, $-S$, and $N$ play the role of momenta, the equation of
state is described by a 3-dimensional surface on which several additional identities, corresponding
to the so-called thermodynamical potentials, hold. This results in the fact that this 3-dimensional
surface is a Lagrangian manifold\footnote{In the data base MathSciNet (American Mathematical
Society) the query to MathJax on Publications results for ``Anywhere =(Maslov index) OR Anywhere
(Maslov class)'' in a list of 575 papers published before 2013, where the notions ``Lagrangian
submanifolds''~\cite{Teor_Vozm}, ``Maslov index'', ``Maslov class'' are developed and extended. See
also \cite{Quantum_Evol}--\cite{Theor_Optics} and so on.}, and the thermodynamical potential
corresponds to action in mechanics.

If one does not consider the number of particles and the volume separately, and only considers their
ratio, i.e., density, then one variable turns out to be redundant, and we can consider 4-dimensional
phase space and the 2-dimensional Lagrangian surface.

If the volume $V$ is given, we can use a single thermodynamical potential
$\Omega$ that can be expressed as follows:
\begin{equation}\label{1s}
d\Omega = (-S)\, dT+  N\,d(-\mu),  \qquad \Omega=-PV .
\end{equation}

Besides, since the internal energy is ${M}=\text{const} \cdot PV$, it is customary in
thermodynamical diagrams to plot the pressure $P$ along the abscissa axis, as this was done in the
Hougen--Watson diagram.

On the $P_r$-$Z$ diagram, the {\it spinodal} is defined as the curve which is the locus of all
points where the tangents to the isotherms are perpendicular to the $P_r$ axis (see
Figures~\ref{mas_000} and \ref{mas_00}).

The mathematical model constructed by the author in the papers~\cite{MTN_94-5}--\cite{RJMP_21-2},
which uses the density ``effect'' mentioned in Section 3 above, shows that

1) if the spinodal in the gaseous region for any gas with non-polarized molecules is given, then all
the gas isotherms can be constructed;

2) if we know the slope of the isotherm for $P_r=0$ on the $P_r$-$Z$ diagram    (this is equivalent
to knowing the value of the second virial coefficient~\cite{Burshtein}) as well as the critical
isochore for any fluid (the supercritical state for any collection of non-polarized molecules), then
we can construct all the isotherms of the fluid up to the critical isochore;

3) if we know the liquid binodal for any gas consisting of non-polarized molecules, and also know
the pressure $P_r$ and the density $\rho$ on the line  $Z=1$ (the Zeno  line), then we can construct
the liquid isotherms  passing through the point $P_r=0$, $Z=0$ for so called ``hard'' liquid
consisting of non-polarized molecules on the interval $0\leq Z\leq 1$.

For an arbitrary liquid, by the {\it Temperley temperature} we mean the minimal temperature of the
set of liquid isotherms passing through the point $P_r=0$, $Z=0$. This temperature (corresponding to
$K= \infty$ ), calculated by the physicist Temperley for the Van-der-Waals gas, equals $3^3/2^5
T_c$~\cite{Temperly}.

As proved by the author in~\cite{MTN_94-5}--\cite{RJMP_21-2}, the corresponding distribution can be
expressed in terms of the polylogarithm\footnote{Note that for $z=1$ and $Re(s)>1$,
$Li_s(1)=\zeta(1)$ is Riemann's zeta function.}
\begin{equation}\label{Li-0}
\operatorname{Li}_{s}(z) = \sum_{k=1}^\infty \frac{z^k}{k^s} = z +\frac{z^2}{2^s}+\frac{z^3}{3^s} + \dots
\end{equation}
This distribution yields the following expressions for the thermodynamical
potential $\Omega$, the density (concentration) $\rho$ and the reduced pressure $P_r$:
\begin{equation}\label{omega2a}
\Omega= - {\Lambda}(\gamma, K) T_r^{2+\gamma}
\bigg\{\operatorname{Li}_{2+\gamma} (a) -
\frac{1}{(K+1)^{1+\gamma}}\operatorname{Li}_{2+\gamma} (a^{K+1})\bigg\},
\qquad \gamma=\gamma(T_r).
\end{equation}
\begin{equation}
\rho(T_r,\gamma,K)
=\frac{\Lambda(\gamma, K)}{\zeta(2+\gamma_c)} T_r^{1+\gamma}
\bigg(\operatorname{Li}_{1+\gamma}(a) - \frac{1}{(K+1)^\gamma}  \operatorname{Li}_{1+\gamma}(a^{K+1})\bigg),
\label{eq-1}
\end{equation}

\begin{equation}
P_r(T_r,\gamma,K)
= \frac{\Lambda(\gamma, K)}{\zeta(2+\gamma_c)} T_r^{2+\gamma}
\bigg(\operatorname{Li}_{2+\gamma}(a) - \frac{1}{(K+1)^{1+\gamma}}  \operatorname{Li}_{2+\gamma}(a^{K+1})\bigg),
\label{eq-2}
\end{equation}
where $K$ is the maximal occupation number, $a=e^{\mu/T_r}$ ($a$ is called the activity),  $\gamma =
D/2-1$,  $D$ is the collective number of degrees of freedom,   $\Lambda(\gamma, K)$  is a
normalizing constant.

For each type of gas, the critical parameters $T_c$, $P_c$, $\rho_c$ are determined experimentally.
Substituting these values into~\eqref{eq-1}--\eqref{eq-2}, we can determine the parameters
$\gamma=\gamma_c$ and $K=K_c$. The parameter $\gamma_c$ is the number of collective degrees of
freedom $D_c= 2\gamma_c +2$ of the critical isotherm. By $K_c$ we denote the admissible size of
clusters (maximal occupation number)\footnote{The separation of the clusters of molecules near the
flat surface of a liquid for $T=T_c$ can be observed experimentally.} at the energy level of the
critical point.

Using~\eqref{omega2a} and the fact that
$$
{M}=-\mu(\partial\Omega)/(\partial\mu) - T(\partial\Omega)/(\partial T) +\Omega,
$$
where $\Omega=-PV$, we derive
${M}=(D/2)PV$.

\section{Entropy and Negentropy}

Above we introduced the notion

 Let
$l$~be the number of particles on a given energy level.
 If we
impose the restriction that each energy level can host at most~$K$
particles,
$l\leq K$,
then we arrive at the Gentile statistics,
also known as parastatistics, and we have the following relation,
which generalizes the Bose--Einstein statistics as
$K\to \infty$
and the Fermi--Dirac statistics for
$K=1$:

\begin{align}
\label{6}
N&=\frac{\Lambda}{\Gamma(\gamma+1)}
\int_0^\infty\bigg(\frac{1}{(e^{\xi/T-\mu/T}-1)}-\frac{K+1}{(e^{(K+1)(\xi/T-\mu/T)}-1)}\bigg)  \xi^\gamma \,d\xi,
\\
\label{61}
M&=\frac{\Lambda}{\Gamma(\gamma+2)}
\int_0^\infty\bigg(\frac{1}{(e^{\xi/T-\mu/T}-1)}-\frac{K+1}{(e^{(K+1)(\xi/T-\mu/T)}-1)}\bigg)  \xi^{\gamma+1} \,d\xi,
\end{align}
where where $T$ is the temperature,  $\gamma =D/2-1$, and $\Lambda$ is the parameter to be
determined in number theory.
 In quantum mechanics, it depends in a well-known way on the mass
and the Planck constant~\cite{MTN_102-2}.

 Indeed, if
$K=1$,
then (without restrictions on the value of the
chemical potential~$\mu$) we obtain the Fermi distribution
\begin{equation}\label{7}
N|_{K=1}=\frac{\Lambda}{\Gamma(\gamma+1)} \int_0^\infty\frac{1}{\exp\{(\xi-\mu)/{T}\}+1}\mspace{2mu} \xi^\gamma \,d\xi.
\end{equation}

 For
$K=\infty$,
the formula coincides with the usual Bose--Einstein
distribution.
 For
$\mu=0$
and
$\gamma=0$
the solution has a
singularity; it becomes infinite.

 To regularize the integral~\eqref{6} for
$K=\infty$,
since
$l\leq N$,
we apply formula~\eqref{6} with
$K= N$.
 From~\eqref{6}, we obtain the
relation
\begin{equation}\label{8}
N = \Lambda \int_0^\infty \bigg(\frac{1}{\exp\{{\varepsilon}/{T}\}-1} -
\frac{N+1}{\exp\{(N+1){\varepsilon}/{T}\}-1}\mspace{2mu}\bigg)\, d\varepsilon.
\end{equation}

Since the relations
\begin{equation}\label{10}
N=-\Lambda T \log(1-a), \qquad
M=\Lambda T^2 \operatorname{dilog}(1-a),
\end{equation}
where $\operatorname{dilog}(1-a)= Li_2(a)$,
hold for the Bose--Einstein distribution with
$a<1$
in the
two-dimensional case, we start by finding~$\Lambda$.

 Since the Ramanujan formula for
$\bar{p}_M(M)$
gives the first term
of the expansion as
$M\to \infty$
in the form
\begin{equation}
\label{12}
\log_e \bar{p}_M(M) \cong2\sqrt{{M}{\operatorname{dilog}(0)}}\,,
\end{equation}
and the first term of the expansion of the entropy is of the form (see \cite[formula~20]{Auluck})
\begin{equation}\label{Sp}
S\cong2\Lambda \sqrt{{M}{\operatorname{dilog}(0)}}\,,
\end{equation}
we see that
$\Lambda\equiv 1$.

 Let $N=N_c$ be the solution of Eq.~\eqref{8}.
Consider the value of the integral  in~\eqref{8} (with the same integrand) taken from
$\delta$
to
$\infty$
and then pass to the limit as
$\delta \to 0$.
 After making the change
$\beta x = \xi$
in the first term and
$\beta (N_c+1)x = \xi$
in the second term,  where $\beta=1/T$,
we obtain
\begin{align}
N_c&=\frac1\beta\int_{\delta \beta}^\infty\frac{\,d\xi}{e^\xi-1}
-\int^\infty_{\delta \beta (N_c+1)}\frac{\,d\xi}{e^\xi-1} +O(\delta)=\frac1\beta
\int^{\delta \beta (N_c+1)}_{\delta \beta}\frac{\,d\xi}{e^\xi-1}+O(\delta)
\label{12a}\\
&\sim \frac1\beta\int^{\delta \beta (N_c+1)}_{\delta
    \beta}\frac{\,d\xi}\xi+O(\delta) =\frac1\beta\{\ln(\delta
\beta (N_c+1))-\ln(\delta \beta)\}+O(\delta)=\frac 1\beta\ln (N_c+1) +O(\delta). \label{13a}
\end{align}
On the other hand, after neglecting the second summand in the integrand of \eqref{61}
for large $M$ and making the change
$\beta x=\xi$,
we can write
\begin{equation}\label{t_m}
\frac1{\beta^2}\int^\infty_0\frac{\xi \,d\xi}{e^\xi-1}\cong M.
 \end{equation}
After passing to the limit as $\delta\to0$ and expressing $\beta$, from~\eqref{t_m}, we obtain
 \begin{equation}\label{11}
N_c= \sqrt{\frac{M}{\operatorname{dilog}(0)}}\mspace{2mu} \log (N_c +1).
 \end{equation}

 The ideal entropy is the same for the Fermi and Bose particles
(see~\cite[Eq.~(13a) as well as Eqs.~(20) and (27)]{Auluck}).

 Hardy and Ramanujan derived the formula
\begin{equation}\label{p_m}
p(M) =\frac{1}{(4\sqrt{3})M} e^{\pi \sqrt{\frac{2M}{3}}}.
\end{equation}

 The asymptotic formula for~$q(M)$
has the form~\cite{Abram}
\begin{equation}\label{q_m}
q(M) =\frac{1}{(4\cdot3^{1/4}) M^{3/4}} e^{\pi \sqrt{M/3}}.
\end{equation}

 If a fermion nucleus emits a single neutron, it becomes a boson,
and vice versa.
 Thus, the energy density flattens out under the
successive emission of neutrons according to Bohr's
concept~\cite{Bohr_Kal}.

 By~\eqref{p_m} and \eqref{q_m}, boson fission overtakes  fermion
fission as $M\to \infty$.
 For a boson not to become a fermion when
emitting neutrons, it must emit them in pairs, which may bring
about fission of the nucleus.
 For the Bose--Einstein distribution~\eqref{11}, the critical value of~$N_c$ satisfies the implicit relation
\begin{equation}\label{n2}
 N_c= \sqrt{\frac{M}{\zeta(2)}}\mspace{2mu}\ln (N_c +1).
\end{equation}

 Therefore,
\begin{equation}\label{n2a}
M= \zeta(2) \bigg(\frac{N_c}{\ln N_c}\bigg)^2.
\end{equation}

As we see, this value corresponds to the limit value of activity $a=1$
(i.e., the chemical potential $\mu=0$).
In this case, we use the Gentile statistics to obtain self-consistent equations.
Similarly, for the Fermi system, this limit values corresponds to $a \to -\infty$ so that
\begin{equation}\label{n2b}
M=  \frac{N_c(N_c-1)}{2} +
 \zeta(2) \bigg(\frac{N_c}{\ln N_c}\bigg)^2.
\end{equation}
These relations can be derived using the Bose and Fermi statistics and the Gentile statistics.

By the Bohr--Kalckar ``correspondence principle''
between the physical notion of nucleus and number theory,
we can transfer these relations to the above-cited construction of number theory
with the zeros taken into account~\cite{MTN_102-2},~\cite{RJMP_24-3}.
We have
\begin{equation}\label{n2c}
p\bigg(\zeta(2) \bigg(\frac{N_c}{\ln N_c}\bigg)^2\bigg) =
q\bigg(\frac{N_c(N_c-1)}{2} +
 \zeta(2) \bigg(\frac{N_c}{\ln N_c}\bigg)^2 \bigg).
\end{equation}
This formula allows us to determine the point $N_c$ such that, for $N>N_c$,
the entropy of the branch with repeated terms and zeros
becomes greater than the entropy of the branch without repetitions and zeros.

\begin{figure}[h!]
\centering
\includegraphics[draft=false]{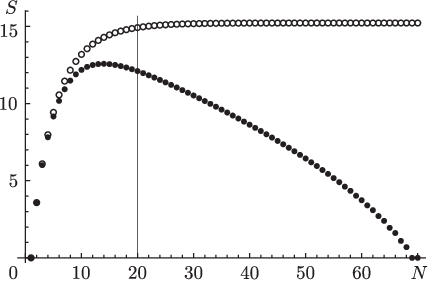}
\caption{The dependence of the entropy $S$ on $N$ for $M=70$.
 The upper curve is obtained when zero summands are allowed, the lower one, when they are not.
 The vertical line indicates the value $N = N_c$ calculated according to formula~\eqref{n2}.
 The exact value of the first point of the phase transition is the maximum of the lower curve.}
\label{mas_06}
\end{figure}

The mesoscopic values in number theory correspond to dimension (number of degrees of freedom) equal to~$2$,
i.e.  in the region, where the parameter $k$  varies from~0 to~90,
Let us find these values $N_c$ from the relation
 \begin{equation}\label{19}
M=T^2\zeta(2) \bigg(1-\frac{1}{(N_c+1)}\bigg),
 \end{equation}
where $T$ can be found from the relation~\cite{Econ_arxiv}
 \begin{equation}\label{20}
 N_c=T\ln (N_c+1).
 \end{equation}
These relations provide an exact dependence of $N_c$ on~$M$.
The graph for $M=70$ is shown in Fig.~\ref{mas_06}.
For $M=70$, the value of $N_c$ is 20. The  value of the entropy~$S$ is 15.
The entropy does not practically vary up to $M=70$, i.e., up to $N=M$.

The point of maximum of the lower curve in
Fig.~\ref{mas_06} for number theory is determined
precisely. This is the value~$N_{\max}$ for which the number $p_N(M)$
attains its maximum.

Now let us return to thermodynamics.

\newpage

\section{The Notion of Generalized Ideal Gas}\label{gen_id-gas}

We pass to the definition of the new classical ideal gas.

Note that in the description of the behavior of multi-particle systems it is possible that the
initial conditions involve interactions. Consider the following analogy. There is a battery of
canons. The gunners position the canons and interact with each other in the process. At some moment
the command ``Fire!'' is heard (it means ``all interactions (all conversations) must stop''). At the
initial situation interactions took place, but once the canon balls are fired, they  fly without
interacting with each other. However elastic collisions between the balls are possible.

This model can be regarded as a model of ideal gas in the traditional understanding only if the
initial data is given at time ``minus infinity''.

\emph{A generalized classical ideal gas is a classical multi-particle system, where interactions
cannot occur only inside certain open regions of the diagram, although at the initial moment of time
or on the boundary of a region interactions between particles are allowed. In the region itself,
only elastic collisions between particles are possible.}

Why is it important to specify the open regions where no interaction (more precisely,  no
attraction) occurs? Obviously, if one must rapidly mix ball bearings of different sizes, it is
easier to do this if the balls are not magnetized. The solution of this problem has several
practical applications. Thus the speed with which two gases mix influences the driving force of jet
engines: the higher the speed, the higher the driving force. If there is no attraction between
molecules of a mixture of gases moving in a pipeline, the probability of the appearance of liquid
bottlenecks is considerably less.

Thus the parameters of the Van-der-Waals gas coincide closely enough with the parameters of the new
(generalized) gas in the following open regions of the parameters: 1) the region from the critical
isotherm  ($T_r=1$) to the critical isochore ($\rho=\rho_c$); 2) the region outside the gas
spinodal; 3) the region of hard liquid (below the Temperley temperature, see Section~\ref{Tem}).

\section{Comparison of the Generalized Ideal Gas with the Van-der-Waals model}~\label{Tem}

The main practical results that follow from the author's theory are related  to metastable states.
In that sense, the main notion of the thermodynamics described here is not the binodal: the curve on
the diagram that indicates the stable phase transition ``vapor-to-liquid'' (the gas binodal) and the
phase transition ``liquid-to-vapor'' (the liquid binodal ). The main notion is the spinodal -- the
curve on the diagram that delimits the regions of stable and metastable states (a more precise
definition of the spinodal was given above). It is practically impossible to reach values of the
thermodynamical parameters on the spinodal experimentally, because that requires ``infinite time''.

From relations~\eqref{eq-1}--\eqref{eq-2} as $K\to\infty$ we obtain the following equations:
\begin{equation}\label{eq-3}
P_r=\frac{\Lambda (\gamma)T_r^{2+\gamma} \operatorname{Li}_{2+\gamma}(a)}{\zeta(2+\gamma_c)},
\qquad a=e^{\mu/T_r},
\qquad \Lambda(\gamma)=\Lambda(\gamma, \infty)=\lim_{K\to\infty}\Lambda(\gamma, K),
\end{equation}
\begin{equation}\label{eq-4}
\rho=\frac{\Lambda(\gamma) T_r^{1+\gamma} \operatorname{Li}_{1+\gamma}(a)}{\zeta(2+\gamma_c)}.
\end{equation}

For $a=1$, we obtain the spinodal defined above. This fact was proved by the author in~\cite{MTN_97-3}.

The constants $\Lambda$ and $\gamma$ can be found for $a=1$  ($\mu=0$) from the coincidence of the
values of  $P_r$ and $\rho$ from~\eqref{eq-3} and~\eqref{eq-4} with the values of $P_{\text{sp}}$
and $\rho_{\text{sp}}$ on the Van-der-Waals spinodal.

\begin{figure}[h!]
\centering
\includegraphics[draft=false]{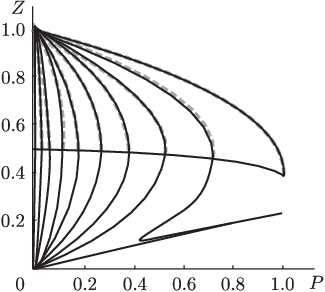}
\caption{On this diagram the pressure is  $P=P_r$. Coincidence of the Van-der-Waals isotherms (solid
lines) with the isotherms from the author's model (dashed lines). The following isotherms are shown
$T_r = 0.1$, $T_r = 0.2$, $T_r = 0.3$, $T_r = 0.4$, $T_r = 0.5$, $T_r = 0.6$, $T_r = 0.7$, $T_r =
0.8$, $T_r = 0.9$,  $T_r = 1$. The value of $\gamma=\gamma(T_r)$ in equations
\eqref{eq-3}--\eqref{eq-4}  is determined from the relations on the Van-der-Waals spinodal
$Z=Z_{sp}$,  $P=P_{sp}$ for $a=1$.} \label{mas_01}
\end{figure}

\begin{figure}[h!]
\centering
\includegraphics[draft=false]{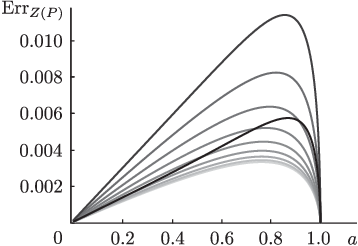}
\caption{Discrepancy between the isotherms of the Van-der-Waals model and the isotherms obtained
from the new distribution;  $a=e^{\mu/T_r}$ is the activity,  $T_r$ is the reduced temperature. The
maximal discrepancy is  $\sim 0.01$  for  $T_r=0.9$ (the effect of $K\neq \infty$ is substantial).}
\label{mas_02}
\end{figure}

In Figures~\ref{mas_01} and \ref{mas_02}, the isotherms of the Van-der-Waals model are compared with
those of the author's model on Hougen--Watson diagrams.

In Figure~\ref{mas_02}, it is shown how the isotherms of the ideal gas (in the new sense) obtained
from ~\eqref{eq-3}--\eqref{eq-4}, differ from the isotherms of the Van-der-Waals model. The figure
shows that the isotherms practically coincide except at points near critical values. In regions with
less than critical parameter values, the coincidence is up to 0.006\%, while at points near critical
values the error is of the order of 0.01\%. Note the important point  $0.9T_c$.  At that point, the
maximal discrepancy between the isotherms is observed. It is of the order of $\sim 0.01$.

For isotherms corresponding to temperatures less than the Temperley temperature (``cluster-free
liquids''), we obtain a very precise coincidence between the isotherms constructed according to the
distribution~\eqref{eq-3}--\eqref{eq-4} and the isotherms coming from the Van-der-Waals model. The
temperature $T_{\text{Temperly}}$ specifies a boundary above which the number $K$ (maximal
occupation number) becomes infinite.

\section{Analytical number theory and the energy of transition
of the Bose gas to the Fermi gas}

Niels Bohr~\cite{Bohr_Kal} notes an important relationship between the number theory
and the entropy, which arises in his concept of the nucleus.
After that paper, several physicists, especially Hindu physicists~\cite{Auluck}--\cite{Rovenchak},
studied this problem and calculated the entropy for the Bose--Einstein gas
and the entropy corresponding to the Ramanujan formula.
These studies have been carried out up to now.

We show that, in the two-dimensional case,
the formulas that were written for the Bose--Einstein distribution in the book~\cite{Landau_StPhys}
coincide with the formulas of analytical number theory.

Indeed, assume that there is a decomposition
$$
M= a_1+ \cdots + a_N
$$
of a number $M$ into $N$ terms. By $N_j$ we denote the number of terms
exactly equal to the number $j$
in the right-hand side of this decomposition.

Then the total number of terms is $\sum_j N_j$, and this number is equal to $N$,
because we known that there is $N$ terms at all.
Further, the sum of terms equal to $j$ is equal to $jN_j$,
because there are $N_j$ of them and the sum of all terms is then obtained
by summing these expressions over $j$, i.e.,  $\sum_j j N_j$, and it is equal to $M$.
Namely,
\begin{equation}\label{1x}
\sum_{i=1}^\infty  N_i=N, \qquad  \sum_{i=1}^\infty  i N_i=M.
\end{equation}

Let us consider an example of the well-known Erd\H{o}s theorem in the number theory,
i.e., the solution of ancient problem called ``partitio numerorum'' in Latin.
This problem deals with an integer $M$ which is decomposed into $N$ terms, for example, $M=5$, $N=2$:
$$
 5=1+4=2+3=0+5,
$$
which gives three versions $\mathcal{M}$ of the solution of this problem: $\mathcal{M}=3$.

If $M=10^{23}$, $N=1$, then there is only one version: $\mathcal{M}=1$.
If $M=10^{23}$ and $N=10^{23}$, then there is also only one version, i.e.,
the sum of units, $\mathcal{M}=1$.

Obviously, for a fixed $M$, there exists a number $\tilde{N}$
for which the number of decomposition versions $\mathcal{M}$ is maximal
(in general, this number is not unique).
The number $\log_2\mathcal{M}$ is called the Hartley entropy.
At the point where it attains its maximum, the entropy is also maximal.
The chemical potential equal to zero corresponds to this point.
The Erd\H{o}s formula determines the maximal number of solutions
of the decomposition function $\tilde{N}$ and has the form
\begin{equation}\label{Erd}
\tilde{N}=\beta^{-1}M ^{1/2} \log
M + \alpha M ^{1/2}+o(M ^{1/2}), \qquad \beta=\pi\sqrt{2/3},
\end{equation}
where the coefficient $\alpha$ is determined by the formula $\beta/2=e^{-\alpha\beta/2}$.

Erd\H{o}s obtained his result only up to $o(\sqrt{M})$
because of the nonuniqueness of the above-mentioned maximum
and the ambiguity of the number of these maxima

\begin{remark}
The classical number theory deals with the space $\mathbb{Z}$ of integers.
Generalizations to the space~$\mathbb{R}^+$ were obtained in the works
which were surveyed in detail by A.G.~Postnikov in the book~\cite{Postnikov}.
This approach agrees well with the concept of statistical physics,
where some of the variables belong to~$\mathbb{R}$,
and hence the integers such as the number of particles~$N$
and the parastatistical number $k$ (restricting the number of particles at one energy level),
together with various numbers belonging to the set~$\mathbb{R}$,
loose their meaning of integer numbers.
\end{remark}

Erd\H{o}s considered the case where the number $N$ is fixed.
We shall consider the sum of all decomposition versions less than or equal to~$N$.
Obviously, in this case, the maximal entropy is also between~$N$ and zero.
The formula is of the same order as in the Erd\H{o}s,
but the answer is significantly different.

Let us consider this question in more detail.

The following formulas hold
for the Bose--Einstein distribution in the case of $D$ degrees of freedom~\cite{Landau_StPhys},~\cite{Abram}:
\begin{equation}\label{BE}
E=\Phi T (\gamma+1)\operatorname{Li}_{2+\gamma}(a),\quad N=\Phi \operatorname{Li}_{1+\gamma}(a).
\end{equation}
Here and below, $\gamma=D/2-1$, $T$ is the temperature,
$a=e^{\mu/T}$ is the activity, $\mu$ is the chemical potential.

The following formulas hold for the Fermi--Dirac distribution:
\begin{equation}
\label{FD}
E=-\Phi T(\gamma+1)\operatorname{Li}_{2+\gamma}(-a),\ N=-\Phi \operatorname{Li}_{1+\gamma}(-a).
\end{equation}
We see that formulas~\eqref{BE} and~\eqref{FD} differ in sign, and hence the activity~$a$
passes through the point $a=0$.

\section{Boson--Ferminon Transition in Mesoscopy}

Our goal is to determine the energy of transition of Bose particles to Fermi particles.
For this, it suffices to study the transition of~\eqref{BE} into~\eqref{FD} through the point $a=0$.
If we want to extend the interval of the jump from the Bose distribution to
the Fermi distribution, then it is necessary to use the parastatistics or the Gentile statistics~\cite{Gentile}.

In the case of parastatistics, we have relations, where the first term in parentheses
gives the distribution for Bose particles, and the second term, the parastatistical correction:
\begin{equation}\label{ep}
E={\Phi}T({\gamma+1}) (\operatorname{Li}_{2+\gamma}(a)-\frac{1}{(k+1)^{\gamma+1}}\operatorname{Li}_{2+\gamma}(a^{k+1})),
\end{equation}
\begin{equation}\label{np}
N= {\Phi} (\operatorname{Li}_{1+\gamma}(a)-\frac{1}{(k+1)^{\gamma}}\operatorname{Li}_{1+\gamma}(a^{k+1})).
\end{equation}

We let $N_i$ denote the number of particles at the $i$th energy level.
In the case of parastatistics, there can be at most $k$ particles at each energy level.
By the usual definitions, the Fermi case is realized for $k=1$, and the Bose case, for $k=\infty$.
But by formulas~\eqref{1x}, it is obvious that $N_i\leq N$ for the Bose system.
Therefore,  $k\leq N$ for the Bose system.
This implies that the maximal value of $k$ is equal to $N$, but not to infinity.
This logical conclusion significantly changes the formulas.
In particular, there arise new equations for $N=k$:
\begin{equation} \label{epN}
E={\Phi}T({\gamma+1}) (\operatorname{Li}_{2+\gamma}(a)-\frac{1}{(N+1)^{\gamma+1}}\operatorname{Li}_{2+\gamma}(a^{N+1})),
\end{equation}
\begin{equation} \label{npN}
N= {\Phi} (\operatorname{Li}_{1+\gamma}(a)-\frac{1}{(N+1)^{\gamma}}\operatorname{Li}_{1+\gamma}(a^{N+1})).
\end{equation}

The maximal number of particles at the energy level $N_c$ in the system
occurs at the caustic point\footnote{In thermodynamics, this caustic is called a spinodal.} $a=1$.

\begin{figure}
\centering
\includegraphics[width=0.7\linewidth]{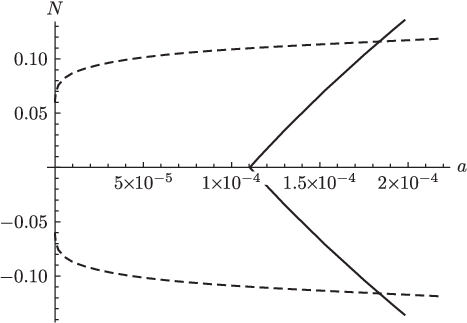}
\caption{Graph of the dependence $N(a)$ for $W=1000$, $W=V(\lambda^2 T)^{\gamma+1}$,
where $\lambda$ is a parameter depending on the mass and $\gamma=0$.
The dotted curve corresponds to $N=-1/\log(a)$.
The solid line corresponds to $N=0$.}
\label{masl_fig_01}
\end{figure}

\subsection{Case of small N}

As was pointed out above, the author obtained a self-consistent equation~\eqref{npN}, where~$N$ is an unknown quantity.
We are interested in the point at which the number of Bose particles $N$ is equal to~0,
i.e., the point at which the Bose particles disappear.
Figure~\ref{masl_fig_01} illustrates an example of this point.
We stress that the value of the activity $a$ does not vanish at this point,
but significantly depends on the function $\Phi$ and the parameter~$\gamma$.

Let us consider the mesoscopic case.
We expand the right-hand side of Eq.~\eqref{npN}
 in a power series in $N\to 0 $.
Using the identity for the polylogarithm
\begin{equation}\label{pl}
z\frac{\partial \operatorname{Li}_s(z)}{\partial z} =\operatorname{Li}_{s-1}(z),
\end{equation}
we  obtain the following expansion in the Taylor series in a small~$N$:
\begin{gather}
\frac{1}{(N+1)^{\gamma}}=1-\gamma  N+\frac{1}{2} \left(\gamma ^2+\gamma \right) N^2+O\big(N^3 \log^3(a)\big),
\nonumber \\
\operatorname{Li}_{1+\gamma}(a^{N+1})=\operatorname{Li}_{\gamma+1}(a)+N \log (a)
\operatorname{Li}_\gamma(a)+\frac{1}{2} N^2 \log ^2(a)\operatorname{Li}_{\gamma-1}(a)+O\big(N^3\log^3(a)\big).
\nonumber
\end{gather}

This implies
\begin{equation}\begin{split}
&\operatorname{Li}_{1+\gamma}(a)-\frac{1}{(N+1)^{\gamma}}\operatorname{Li}_{1+\gamma}(a^{N+1})
=N (\gamma \operatorname{Li}_{\gamma +1}(a)-\log (a)\operatorname{Li}_{\gamma }(a))
\\
&+\frac{1}{2} N^2 \left(\gamma ^2 (-\operatorname{Li}_{\gamma +1}(a))-\gamma \operatorname{Li}_{\gamma +1}(a)
-\log ^2(a)\operatorname{Li}_{\gamma -1}(a)+2 \gamma  \log (a)\operatorname{Li}_{\gamma }(a)\right)
+O\big(N^3 \log^3(a)\big).
\end{split}
\end{equation}

Then the expansion of the number $N$ becomes (up to  $o(N^2\log^2(a))$):
\begin{equation} \label{Mas1a}
\begin{split}
&\frac{N}{V}=(\lambda^2 T)^{\gamma+1}[N (\gamma \operatorname{Li}_{\gamma +1}(a)-\log (a)\operatorname{Li}_{\gamma }(a))
\\
&+\frac{1}{2} N^2 \left(\gamma ^2 (-\operatorname{Li}_{\gamma +1}(a))-\gamma \operatorname{Li}_{\gamma +1}(a)
-\log ^2(a)\operatorname{Li}_{\gamma -1}(a)+2 \gamma  \log (a)\operatorname{Li}_{\gamma }(a)\right)].
\end{split}
\end{equation}

Dividing Eq.~\eqref{Mas1a} by $N$, we obtain
\begin{equation}\label{Mas2}
\begin{split}
&\frac{1}{V}=(\lambda^2 T)^{\gamma+1}[ (\gamma \operatorname{Li}_{\gamma +1}(a)-\log (a)\operatorname{Li}_{\gamma }(a))
\\
&+\frac{1}{2} N \left(-(\gamma ^2+\gamma)\operatorname{Li}_{\gamma +1}(a)-\log ^2(a)\operatorname{Li}_{\gamma -1}(a)
+2 \gamma  \log (a)\operatorname{Li}_{\gamma }(a)\right)].
\end{split}
\end{equation}

This implies
\begin{equation} \label{eg-2}
\begin{split}
&-1+V(\lambda^2 T)^{\gamma+1} (\gamma  \operatorname{Li}_{\gamma +1}(a)-\log (a) \operatorname{Li}_{\gamma }(a))
\\
&+\frac{V(\lambda^2 T)^{1+\gamma}}{2} N \left(-(\gamma ^2+\gamma) \operatorname{Li}_{\gamma +1}(a)-\log ^2(a)
\operatorname{Li}_{\gamma -1}(a)+2 \gamma  \log (a) \operatorname{Li}_{\gamma }(a)\right) =0.
\end{split}
\end{equation}

To pass from the case $k=0$ to the case $k=\infty$,
it is necessary to use the natural series of numbers.
We will consider this transition below.
We assume that $k=N$ in an infinitely small neighborhood of the resonance~$[N]$,
where $[N]$ is the integer part of the number~$N$.
The set of points infinitely close to the number~$[N]$
is called a monad in the nonstandard analysis or the Leibnitz differential
(see~\cite{Nestandart-1}--\cite{Nestandart-2}, and also see~\cite{Shepin-1}--\cite{Shepin-2}).
We let~$x$ denote the difference $N-[N]$, i.e., $N-[N]=x>0$.
We see the expansion in a power series in~$x$ up to $O(x^2)$,
which implies $N\sim |N|$.

The self-consistent relation for $x$ near the resonance points $[N]$ has the form
\begin{equation}\label{N}
[N]+x= {(T\lambda)^{\gamma+1} V}(\operatorname{Li}_{1+\gamma}(a)
-\frac{1}{([N]+x+1)^{\gamma}}\operatorname{Li}_{1+\gamma}(a^{[N]+x+1})).
\end{equation}

The following thermodynamic formula is known for the pressure $P$:
\begin{equation}\label{M}
P= T{(T\lambda)^{\gamma+1}}(\operatorname{Li}_{2+\gamma}(a)
-\frac{1}{([N]+x+1)^{\gamma+1}}\operatorname{Li}_{2+\gamma}(a^{[N]+x+1})).
\end{equation}

In the Gentile statistics, each $\Phi$ is associated with its own value
of activity~$a$ at which the number of particles is $N=0$.
In this case, the Bose particles disappear. They split into Fermi particles
and the Bose statistics turns into the Fermi statistics.
We denote such a value by~$a_0$.

Let us consider the behavior of the system in the case of distinct values of~$k$.

\subsection{Expansion near $k=0$}

As was shown above, the number of particles at the energy level
cannot exceed $N$.
We can expand the right-hand side of Eq.~\eqref{N} in small~$N$
neglecting the terms $O((N-k)^2)$ as
\begin{equation}     \label{N01}
\begin{split}
&N=\Phi N (\gamma \operatorname{Li}_{\gamma +1}(a)-\log (a)\operatorname{Li}_{\gamma }(a))\\
&+\Phi \frac{1}{2} N^2 \left(\log ^2(a) (-\text{Li}_{\gamma -1}(a))
-\gamma  ((\gamma +1)\operatorname{Li}_{\gamma +1}(a)
-2 \log (a)\operatorname{Li}_{\gamma }(a))\right).
\end{split}
\end{equation}

Passing to the limit in~\eqref{N01} as $N\to0$, we obtain the equation for $a$
for which $N=0$:
\begin{equation}
\label{N=0}
\gamma \operatorname{Li}_{\gamma+1}(a_0)-  \log (a_0)\operatorname{Li}_\gamma(a_0)-\Phi^{-1}=0.
\end{equation}
For a sufficiently large $\Phi$, this equation has a unique solution which we denote by~$a_0$.

In this case, the error is the difference between the noninteger solution~$N$
for $a>a_0$ and the integer $N=0$ for $a=a_0$, i.e., $x=N-0=N$.

\subsection{Expansion near $k=1$}

Let $N=1+x$, where $x$ is small.
The expansion of the right-hand side of~\eqref{N} for small $x$ leads to the equation for~$x$:
\begin{equation}     \label{N1}
1+x=\Phi[\text{Li}_{\gamma +1}(a)-2^{-\gamma }\operatorname{Li}_{\gamma +1}\left(a^2\right)
-2^{-\gamma -1} x \left(2 \log (a)\operatorname{Li}_{\gamma }\left(a^2\right)
-\gamma \operatorname{Li}_{\gamma +1}\left(a^2\right)\right)+O(x^2)].
\end{equation}

The equation for $a$ at which $x=0$ or $N=1$ has the form
\begin{equation}  \label{N=1}
-\Phi \operatorname{Li}_{\gamma +1}\left(a^2\right)+2^{\gamma }
\Phi \operatorname{Li}_{\gamma +1}(a)-2^{\gamma }=0.
\end{equation}
This equation has a unique solution which we denote by $a_1$ following
the notation introduced above.

In this case, the error is the difference between the solution of the equation
for noninteger~$N$ and $a>a_1$ and the solution of the equation
for integer $N=1$ and $a=a_1$, i.e., $x=N-1$.

The pressure expansion~\eqref{M} for small $x$ up to the first order inclusively
has the form
\begin{equation}\label{E1}
P=\frac{\Phi T}{V} [\left(\text{Li}_{\gamma +2}(a)
-2^{-\gamma -1} \operatorname{Li}_{\gamma +2}\left(a^2\right)\right)
+x \left(2^{-\gamma -2} (\gamma +1)\operatorname{Li}_{\gamma +2}\left(a^2\right)
-2^{-\gamma -1} \log(a)\operatorname{Li}_{\gamma +1}\left(a^2\right)\right)].
\end{equation}

\subsection{Expansion near an arbitrary positive integer $k=i$.}

After similar calculations, one can obtain the asymptotic formulas for the error
in determining~$N$ and the pressures~$P$ near $k=2$:
\begin{equation} \label{N22}
 x=N-2,
\end{equation}

\begin{equation}\label{E21}
P=\frac{\Phi T}{ V}[\left(\text{Li}_{\gamma +2}(a)-3^{-\gamma -1}\operatorname{Li}_{\gamma +2}
\left(a^3\right)\right)-3^{-\gamma -2} x
\left(-\gamma \operatorname{Li}_{\gamma +2}\left(a^3\right)
-\text{Li}_{\gamma +2}\left(a^3\right)+3 \log (a)\operatorname{Li}_{\gamma +1}
\left(a^3\right)\right)].
\end{equation}

For an arbitrary positive integer~$i$, we have the relations
\begin{equation}\label{NK2}
x=N-i,
\end{equation}

\begin{equation}\label{EK1}
\begin{split}
&P=\frac{\Phi T}{ V} [\left(\text{Li}_{\gamma +2}(a)-(i+1)^{-\gamma -1}
\operatorname{Li}_{\gamma +2}\left(a^{i+1}\right)\right)\\
&-x (i+1)^{-\gamma -2} \left(-(\gamma+1) \operatorname{Li}_{\gamma +2}
\left(a^{i+1}\right)+(i+1) \log (a)\operatorname{Li}_{\gamma +1}\left(a^{i+1}\right)\right)].\\
\end{split}
\end{equation}

The equation for $a_i$ at which $N=i$ has the form
\begin{equation}\label{Ni}
-\Phi \operatorname{Li}_{\gamma +1}\left(a^{i+1}\right)+(i+1)^{\gamma }
\Phi \operatorname{Li}_{\gamma +1}(a)-i\ (i+1)^{\gamma }=0.
\end{equation}

\begin{figure}
    \centering
    \includegraphics[width=0.7\linewidth]{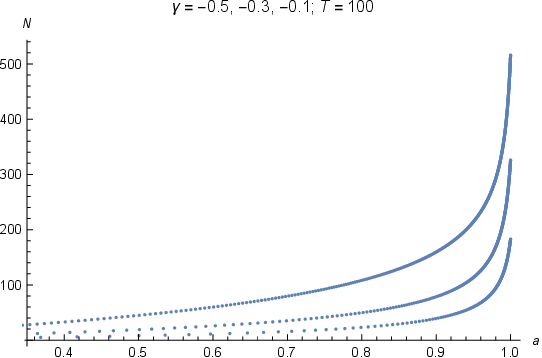}
    \caption{Dependence of the number of particles $N$ on $a$
    for a given temperature $T=100$ which is the result of numerical solution of Eq.~\eqref{Ni}.
    Here  $\gamma=-0.5,-0.3,-0.1$ counting upwards from below.
    For $\gamma=-0.5$, the maximum $N=183$ is attained at $a=0.999906$.
    For $\gamma=-0.3$, the maximum $N=326$ is attained at $a=0.999954$.
    For $\gamma=-0.1$, the maximum $N=516$ is attained at $a=0.999985$.}
    \label{figure_1}
\end{figure}

\begin{figure}
    \centering
    \includegraphics[width=0.7\linewidth]{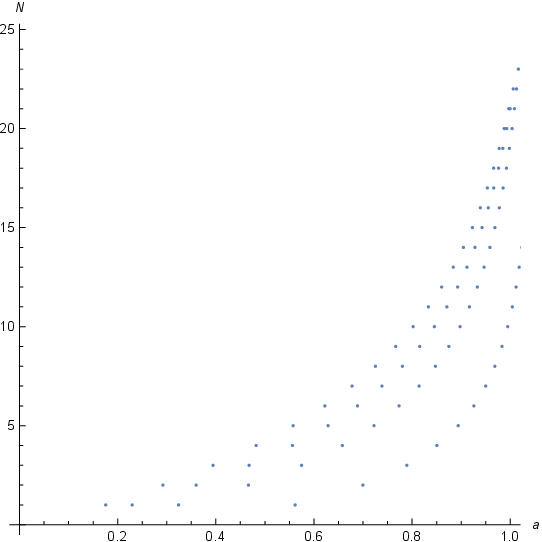}
    \caption{Dependence of the number of particles $N$ on $a$
    for a given energy $M=70$ which is the result of numerical solution
    of the system of Eqs.~\eqref{N},~\eqref{M}.
    Here $\gamma=-0.7,-0.5,-0.3,-0.1$ counting upwards from below.
    The values of maximal $a_i$ are given in~\eqref{t1}.}
    \label{figure_2}
\end{figure}

\begin{figure}
    \centering
    \includegraphics[width=0.7\linewidth]{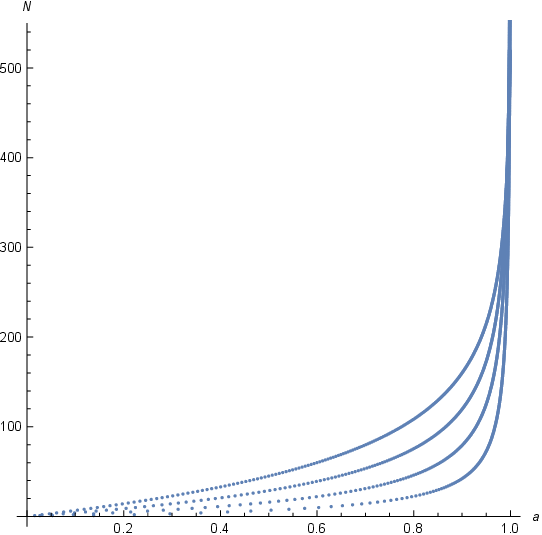}
    \caption{Dependence of the number of particles $N$ on $a$
    for a given energy $M=10000$ which is the result of numerical solution
    of the system of Eqs.~\eqref{N},~\eqref{M}.
    Here $\gamma=-0.7,-0.5,-0.3,-0.1$ counting upwards from below.
    The values of maximal~$a_i$ are given in~\eqref{t2}.}
    \label{figure_3}
\end{figure}

The solutions of Eq.~\eqref{Ni} are shown in Figs.~\ref{figure_1}, \ref{figure_2},
and~\ref{figure_3}.

Note that, for given values of the parameters~$\Phi$ and $\gamma$,
Eq.~\eqref{Ni} has the maximal solution $a_i=a_{max}<1$ corresponding to
the maximal $i=i_{max}$.

As was said above, $M$ is the total energy of all particles.
Putting $\mu=0$, neglecting the negative term in formula~\eqref{M}
as $N\to\infty$, and using the well-known thermodynamical relation
$M=(\gamma+1)PV$, we express the temperature in terms of the parameters
$M,\gamma,\lambda,V$ as follows:
\begin{equation}
\label{TM}
T(M,\gamma,\lambda,V)=
\left(\frac{M}{(\gamma+1)V\lambda^{1+\gamma}\zeta(2+\gamma)}\right)^{\frac{1}{\gamma+2}}.
\end{equation}

Thus, the temperature is determined on the spinodal.

In~\eqref{t1} and \eqref{t2}, we present the corresponding values $a_{max}$, $i_{max}$
for $M=70$ and $M=10000$.

\begin{equation}\label{t1}
M=70:
\left(
\begin{array}{ccc}
\gamma & i_{max} & a_{max} \\
-0.1 & 21 & 0.995525 \\
-0.3 & 22 & 0.991582 \\
-0.5 & 23 & 0.997874 \\
-0.7 & 17 & 0.994192 \\
\end{array}
\right)
\end{equation}

\begin{equation}
\label{t2}
M=10000:
\left(
\begin{array}{ccc}
\gamma & i_{max} & a_{max} \\
-0.1 & 519 & 0.99998 \\
-0.3 & 631 & 0.999991 \\
-0.5 & 784 & 0.999964 \\
-0.7 & 817 & 0.999999 \\
\end{array}
\right)
\end{equation}

\subsection{Jump of the specific energy}

Now we consider the situation where only the parameter~$a$ varies at a fixed temperature,
namely, for the Bose case, it varies from $a_{max}$ to $a_0$,
and for the Fermi case, from $0$ to infinity.

If we consider only traditional formulas for the Bose gas which correspond
to the case where infinitely many particles can be located at each energy level,
then the negative term in the integrand in formulas~\eqref{N} and~\eqref{M}
is absent.
In this case, the transition of a boson particle to a fermion particle
is accompanied by the change of the sign of the activity~$a$
in all formulas, where the polylogarithm function is contained.
Indeed, for the Bose--Einstein distribution, we have
\begin{equation}
\label{LiB}
\operatorname{Li}_{\gamma +1}(a)=\frac{1}{\Gamma{(\gamma+1)}}
\int_0^{\infty}{\frac{\xi^{\gamma}d\xi}{e^{\xi}/a-1}},
\end{equation}
and for the Fermi--Dirac distribution, we have
\begin{equation}
\label{LiF}
-\operatorname{Li}_{\gamma +1}(-a)=\frac{1}{\Gamma{(\gamma+1)}}\int_0^{\infty}{\frac{\xi^{\gamma}d\xi}{e^{\xi}/a+1}}.
\end{equation}
Thus, we can assume that the activity is equal to $-a$ and negative for the Fermi case,
and it is equal to~$a$ and remains positive for the Bose case.

If the activity $a$ changes sign, this means that it necessarily passes
through the transition point $a=0$ at which both the pressure and the number of particles are equal
to zero.
To avoid the change of sign of the polylogarithm function, we assume that the values
of the energy~$M$ and the pressure $P$ and hence the number of particles~$N$
of the Bose gas are negative.

But in contrast to Eqs.~\eqref{N} and~\eqref{M}, the above obtained equations
determine the transition point not for $a=0$ but for $a=a_0>0$.
Precisely at this point, as will be shown below,
the compressibility factor $Z=\frac{P V}{N T}$ and hence the specific energy
experience a jump.

\begin{figure}
    \centering
    \includegraphics[width=0.8\linewidth]{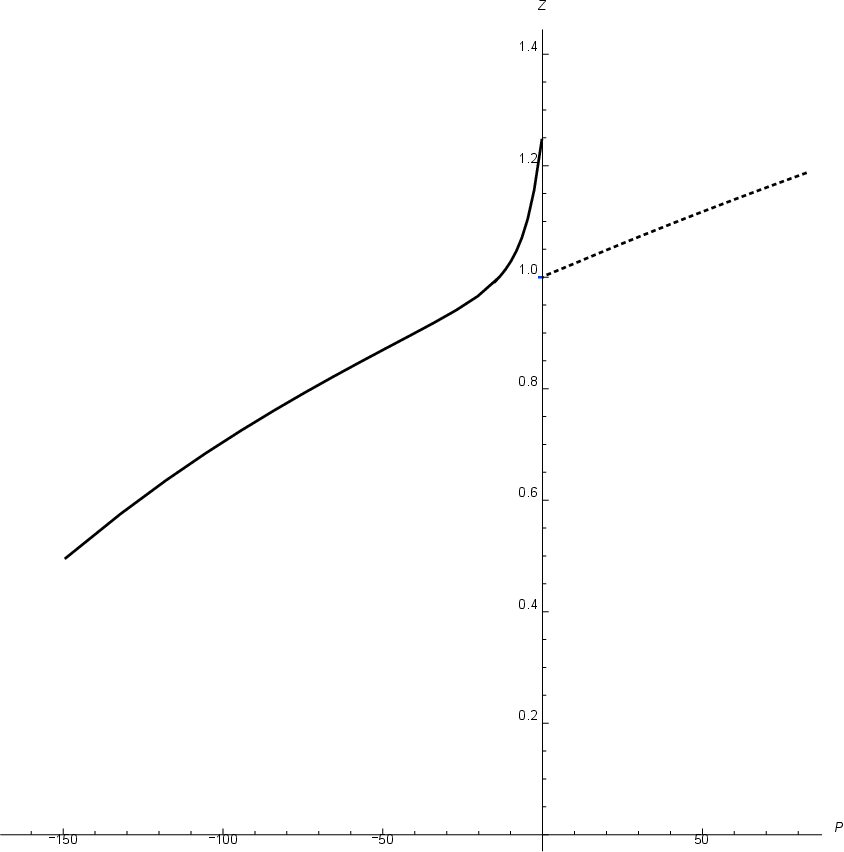}
    \caption{Dependence of the compressibility factor~$Z$ on $P$
    at a given temperature $T=10$.
    The solid curve on the left corresponds to the Bose case,
    the dashed curve on the right, to the Fermi case.
    The jump of the compressibility factor is equal to $\Delta Z=-0.25$.}
    \label{figure_4}
 \end{figure}

The jump $Z$ is visually illustrated in Fig.~\ref{figure_4}.

Starting from the value of the compressibility factor $Z=\frac{P V}{N T}$,
we obtain the expression for the specific energy:
$E_{sp}=\frac{E}{N}=Z(\gamma+1)T>0$.
The limit of the function~$Z$ as $P\to+0$ from the right,
which corresponds to the Fermion branch, is equal to~$1$.
The limit of~$Z$ as $P\to-0$ from the left, as was shown in~\cite{RJMP_25-2},
is equal to $Z_0(M,\gamma,\lambda,V)$, where
\begin{equation}
\label{z1}
Z_0(M,\gamma,\lambda,V)=\lambda^{\gamma+1} V [T(M,\gamma,\lambda,V)]^{1+\gamma}
(\gamma  \text{Li}_{\gamma +2}(a_0)+\text{Li}_{\gamma +2}(a_0)-\log (a_0) \text{Li}_{\gamma +1}(a_0))>1.
\end{equation}

In such a transition, the compressibility factor has a jump by the value
$\Delta Z=Z_{Fermi}-Z_{Bose}=1-Z_0<0$.
This corresponds to the jump of the specific energy equal to
$\Delta E_{sp}=T(\gamma+1)(Z_{Fermi}-Z_{Bose})=T(\gamma+1)(1+Z_0)<0$.

Table~1 presents the values of $Z_0$ for different $\gamma$
for $M=200$ under the condition $\lambda^{\gamma+1} V=1$.
\begin{table}[h!]
    \caption{}
    \tabcolsep=0.3em
    \small
    \begin{tabular}{|c||c|c|c|c|c|c|c}
        \hline
        {$\gamma$} & {-0.7} & {-0.6} & {-0.5}& {-0.4} & {-0.3} & {0}  \\
        \hline
        {$Z_0$} & 1.54 & 1.44 & 1.38 & 1.34 & 1.31 & 1.25 \\
        \hline
    \end{tabular}
\end{table}

\begin{figure}
\centering
\includegraphics[width=0.7\linewidth]{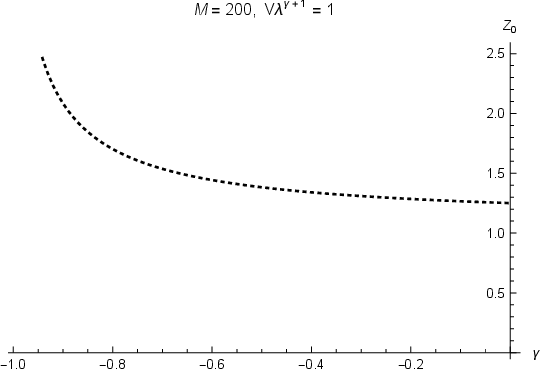}
\caption{Dependence of $Z_0(M,\gamma,\lambda,V)$ on $\gamma$
for a given energy $M=200$, constructed by formula~\eqref{z1}. }
\label{figure_5}
\end{figure}

The dependence of $Z_0(M,\gamma,\lambda,V)$ on $\gamma$ for a given energy $M=200$
is shown in Fig.~\ref{figure_5}, where the behavior of $Z_0$ is shown for the continuously varying value
$\gamma \in (-1,0]$.

\section{Transition of the Helium-6 boson to the Helium-5 fermion}

Following the author's concept related to the abstract analytical number theory~\cite{Postnikov},
one can mathematically calculate the transition of Bose particles to Fermi particles,
at least in the two-dimensional case.
Based on this concept, the boson branch of the decomposition of the number $M$ into terms
(with possible repetition of terms) turns into the fermion branch of the decomposition
(without repeated terms).
It follows from the continuity of such a transition that there exists
a point of transition from the boson branch into the fermion branch
according to the number of terms~$N$.

The specific volume~$V$ (the area in the two-dimensional case)
was determined in  the number theory in~\cite{Masl_Naz_100-3} as $N=V\operatorname{Li}_1(a)$
for the distribution of the Bose gas. Here $N/V$ is the density.
For the volume~$V$, the energy can be written as $E=V \operatorname{Li}_2(a)$.

When the activity $a$ changes the sign, the boson branch
(with repetition of terms in the decomposition of the number) in the number theory
turns into the fermion branch
(without repetition of the terms in the decomposition).
One succeeds in passing from the number theory to the case of small dimension,
i.e., to any number of degrees of freedom greater or smaller than~2.
In this case, on succeeds in calculating the coefficient of $1/\log a$,
which permits determining the value of the energy
required for the Bose gas to go over into the Fermi gas (for example, Helium-4 into Helium-3)
for a given volume and a given temperature.

\subsection{Quantization of  activity  and energy of G\"{o}del,  Maltsev  and Ershov numbering.
 Calculation of the hidden parameter for the microscopy}

In this and other works, the author considered the ``hidden parameter''~$t_{\mathrm{meas}}$, which is not hidden in the
sense of the Einstein--Podolsky--Rosen paradox (EPR). The parameter introduced by the author is quite natural and open.
In the assertion on the identity of particles in the works of Landau and Lifshits~\cite{Landau_Kv_Mech}
which has been many times cited by the author in his papers,
this parameter is veiled as the ``time moment'' at which the numeration of particles is attained.
This citation states the following:
``one can imagine that the particles contained in a given physical system
are ``renumbered'' at a certain time moment''
(p.~252).
The time required to numerate the particles is precisely the additional parameter
introduced by the author in~\cite{RJMP_24-3}, \cite{14}--\cite{15} and which is discussed in the present paper.
This time depends on the algorithm used to numerate the particles.
In turn, the time of implementation of the algorithm depends on the calculating person
and his device. Thus, this parameter is not hidden but is a somewhat veiled.
It can be determined exactly only under a lot of additional conditions.

In the paragraph of the book~\cite{Landau_Kv_Mech} cited above, Landau and Lifshits also speak about other time moments,
namely, if ``one further watches the motion of each of the particles in its trajectory, then the particles
can be identified at any \emph{time moment}'' (italics added by VM).

The time moments form a discrete set of points.
It the intervals between these points are much less than the veiled parameter,
then the observer sees the classical picture of the neutron (wave packet) revolution
around the Helium-4 nucleus independently of the life time of the Helium-5 fermion.

We consider a gas, i.e., a sufficiently many particles, each of which is a boson.
If we consider specific characteristics of the gas (specific energy, specific volume, etc.),
then we can assume that these specific quantities are related to the microscopy,
i.e., to the situation of a single nucleus with several nucleons.

Assume that the hidden parameter $t_{\mathrm{meas}}$
(i.e., the time given to the experimenter for observation)
is of the order of $10^{-11}$, and the life time of a boson is of the order of $10^{-13}$.
Our assumption is not strict, but since the value of time of the order of $10^{-11}$
is much greater than the value of the order of $10^{-13}$,
and hence much greater than the time during which
the experimenter observed 20 boson revolutions around the nucleus,
we can assume that this value of time is approximately equal to the hidden parameter
in the microscopy in the sense explained above.
If we have one nucleus and each nucleon is a boson,
then for the bosons in this approximation (in terms of approximate specific energies),
we can calculate the sum of specific energies for all bosons.
This total energy coincides with the energy of microparticles, i.e. for all nucleons
in the order sufficient for us.

Indeed, we are interested not in the energy but in the hidden parameter $t_{\text{meas}}$.
With regard to the above, we can state the hidden parameter is revealed with a sufficient accuracy,
i.e., with a logarithmic accuracy.

In the present paper, we propose a method for determining the energy of fermions and bosons
in the case of a gas, i.e., in the case where there are very many particles.
For the specific energy, we can add the specific values of separate particles
and obtain the microscopic sum of energies of separate nucleons.
Since the hidden parameter is calculated approximately
(with a logarithmic accuracy), we can justify the method proposed for calculations.

In~\cite{Bohr_Kal}, it is similarly proposed to sum the microscopic energies of particles
to calculate the binding energy of the whole nucleus.
This permits calculating the value of the hidden parameter.

Let us consider the specific energy of a gas consisting of particles obeying
the Bose--Einstein statistics for $a=1$:
\begin{equation}\label{eud}
E_{spec}=T(\gamma+1)\frac{ \zeta(2+\gamma)}{ \zeta(1+\gamma)}.
\end{equation}
This energy corresponds to the macroscopy,
while the specific binding energy of the nucleus $\epsilon$,
i.e., the energy per 1 nucleon, corresponds to the microscopy.
The equality of these energies permits calculating the parameter $\beta=\frac{1}{T}$:
\begin{equation} \label{b}
\beta =(\gamma+1)\frac{ \zeta(2+\gamma)}{\epsilon \zeta(1+\gamma)}.
\end{equation}
The time $t_0$ equal to the interval between the moments of measurement
can be calculated by the formula
\begin{equation} \label{texp}
t_0=\beta \hbar=(\gamma+1)\hbar\frac{ \zeta(2+\gamma)}{\epsilon \zeta(1+\gamma)},
\end{equation}
where $\hbar$ is the Plank constant.

For Helium-5, the specific binding energy is equal to $5.481$\,MeV,
the life time of the nucleus is $1.01\times 10^{-21}$\,s,
and the time $t_0=5.2\times 10^{-22}$\,s for $\gamma=3.5$.
For Helium-6, the specific binding energy is equal to $4.878$\,MeV,
the life time of the nucleus is $1.16$\,d,
and the time $t_0=5.9\times 10^{-22}$\,s for $\gamma=3.5$.

\begin{figure}[h]
\centering
\includegraphics{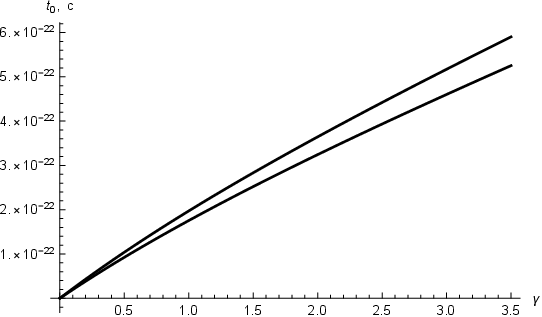}
\caption{Dependence of the time $t_0(\gamma)$ given by formula~\eqref{texp}.
The upper curve corresponds to Helium-6, the lower, to Helium-5.}
\label{masl_fig_02}
\end{figure}

The dependence $t_0(\gamma)$ is depicted in Fig.~\ref{masl_fig_02}.
In particular, for $\gamma=0.5$, the life time for Helium-5
is equal to $9.25\times 10^{-23}$\,S, and for Helimu-6, to $1.04\times 10^{-22}$\,s.

We assume that to distinguish one revolution of a neutron around the nucleus,
the experimenter has to take at least 10 pictures in this time period.
Thus, to distinguish 20 revolutions, it is necessary to take 200
during the time $200 t_0$.
In the case $\gamma=3.5$ for Helium-5, this time is equal to $t_0=1.05\times 10^{-19}$\,s,
which significantly exceeds the time given for the experiment, i.e.,
the value of the hidden parameter.

Thus if the life time of a particle is less that the time given for the experimental observation
(i.e., less that the hidden parameter), then the experimenter can distinguish this particle
from the others. But if the life time of a particle is greater than the observation time,
then the the experimenter cannot distinguish the particles.
But during the time given to him, he can observe some separate moments or stages of the particle life,
for example, 20 revolutions of the particle around the nucleus.
But these separate moments of the particle life do not allow him to distinguish
this particle from the other particles, i.e. to identify it,

In other words, our concept is based on the following three times:
the time given for the experiment (hidden parameter),
the life time of a particle,
and the real time of events in the life of a particle
which is less than the time of its whole life and which is observed by the experimenter.

Since we distinguish the hidden parameter in the quantum mechanics
and the hidden parameter in classical statistical physics,
it is necessary,  more precisely than usually in the physical literature,
to separate the quantum quantities and the classical quantities
which disappear as $\hbar\to 0$.
For example, in the literature,
the light polarization is related to the wave optics,
despite the fact that, as is known,
the polarization does not disappear as the frequency tends to infinity
and hence remains in the geometric optics.

Similarly, the spin does not disappear in the classical limit $\hbar\to 0$
and is preserved in the classical transport equation~\cite{Teor_Vozm},~\cite{16}.
The hidden parameter for the spin is not related to the quantum
hidden parameter $\Delta t=\hbar/\Delta E$.

The WKB method of transition from the quantum mechanics in classical mechanics
leads to the following two classical equations:
the Hamilton--Jacobi equation and the transport equation.
Both of these equations describe the classical mechanics.
The Hamilton--Jacobi equation  contains a term
which determines the interaction of particles in the Newton equation
The interaction between the spin and the magnetic field is contained
in the transport equation, and hence can be explained in classical mechanics.
The question of how to describe the interaction between the terms in the transport equation
if there is no interaction in the Hamilton--Jacobi equation
was posed by Anosov and by the author at the beginning of the 1960s.
At present, this problem is related to the Bell inequality and hence to the hidden parameters
arising in this case~\cite{Bell}.
The author agrees with Bohm in that the main difficulty in revealing the EPR paradox
reduces to a problem related to the spin and polarization.
But as follows from the above, this by no means does not concern the problem
of the quantum hidden parameter considered in this section.
The difficulty is in the problem of spin, and it reduces to the classical mechanics.
Precisely in the same way, we can neglect the Newtonian attraction.

The experiment with rotation of a boson around the nucleus revealed the following.
On the one hand, the experimenter allowed an increase in the hidden parameter to such an extent
that the wave packet corresponding to the nucleon began to differ from its twin.
This implies that the nucleon started to behave not as a quantum particle,
which is located at a separate Bohr energy level,
but as a classical particle,
which rotates as the Rutherford electron.

This fact already shows that, even for small energies,
the nucleon does behave according to the Bohr rule related to separate orbits.
Does this mean that there is a WKB approximation?
The transition to the classics was previously explained only by the WKB approximation.
But the experiments allowed one to calculate the hidden parameter
and confirmed the fact that the hidden parameter is large so that it permits
observing the behavior of a nucleon separately from its twin behavior.
By the hidden parameter law, the nucleon behaves like a wave packet independently of its twin,
and hence it makes revolutions.

Thus, this well-known experimental fact can be explained by using the hidden parameter,
which permits considering the classical mechanics and the quantum mechanics
from the common point of view.
This unifies the two sciences into a comprehensive whole and exactly shows
when we must pass from the quantum theory to the classical theory  without using the WKB method.
This is a principally different method for such a transition as compared to the WKB method.
In what follows, we show the converse; namely, the quantum problem,
in turn, gives important results in classical theory of thermodynamics
without the semiclassical transition.

\section*{Acknowledgments}

The author is deeply grateful to professor V.S. Vorobiev for fruitful discussions as well as for
checking all the diagrams and verifying the author's model on the basis of the latest experimental
results for nitrogen. The author is also grateful to E.P. Loubenets,  S.G. Kadmenskii and A.A. Ogloblin
for useful discussions,  and
D.S. Minenkov    and  E.~I.~Nikulin for help in calculations and construction of graphs.
The author  would like also to express sincere gratitude to A.~B.~Sossinsky for  his help in translation into  English.


\begin{thebibliography}{99}

\bibitem{MTN_94-5}
Maslov V.~P.
Undistinguishing Statistics of Objectively Distinguishable Objects:
Thermodynamics and Superfluidity of Classical Gas.
Math. Notes  2013; \textbf{94} (5):722-813.

\bibitem{RJMP_21-2}
Maslov V.~P.
New Construction of Classical  Thermodynamics and
UD-Statistics.
Russian J.
Math.
Phys.  2014; \textbf{21} (2):256-284.

\bibitem{Elsevier_2014}
Maslov V.~P.  Thermodynamics and Economics: Overview.
Reference Module in Earth Systems and Environmental Sciences. Elsevier; 2014.

\bibitem{Teor_ver_Premia}
Dobrokhotov S.~Yu.,  Karasev M.~V.,  Kolokoltsov V.~N.,  et al.
Award  of the State Prize of the Russian Federation in Science and Technology (2013) to Victor Maslov.
 Probability Theory and Applications  2014; \textbf{59} (2):2009--2013.

\bibitem{Landau_StPhys}
Landau  L.~D.
and Lifshits E.~M.
Statistical Physics.  Moscow: Nauka; 1964
[in Russian].


\bibitem{Kad_Dyn}
Kadomtsev B.~B. Dynamics and Information.
Moscow, Uspekhi Fizicheskikh Nauk; 1999
[in Russian].

\bibitem{Landau_Kv_Mech}
Landau  L.~D.
and Lifshits E.~M. Quantum Mechanics: Non-Relativistic
Theory, 2nd ed.
Moscow:Nauka; 1964 [in Russian];
(translation of the 1st
ed., Pergamon Press, London--Paris
and Addison-Wesley Publishing
Co., Inc., Reading, Mass.; 1958).

\bibitem{Masl-Shv_Com-germ}
Maslov V.~P. and Shvedov O.~Yu.
The Complex Germ Method in Many-Particle Problems and in Quantum Field Theory.
Moscow: Editorial URSS; 2000 [in Russian].

\bibitem{MTN_102-6-shcom}
V.~P.~Maslov,
``Hidden variables in  the theory of measuring,''
Math. Notes \textbf{102} (6),   (2017).

\bibitem{Gentile}
Dai W.-S., Xie M.Gentile
 statistics with a large maximum  occupation number.
 Annals of Physics 2004 \textbf{309}:295--305.

 \bibitem{RJ_21-1}
Maslov V.~P.  The relationship between the Van-der-Waals model and the
undistinguishing statistics of objectively distinguishable objects.
The new parastatistics.
Russian J.
Math.
Phys. 2014; \textbf{21} (1):99--111.

\bibitem{Vershik}
Vershik  A.~ M.
Statistical mechanics of combinatorial partitions, and their limit shapes,''
 Funktsional.
 Anal. i Prilozhen. 1996;
\textbf{30}
(2): 19--39
[Functional Anal.
 Appl.  1996; \textbf{30}
(2):90--105].

\bibitem{MTN_Mas_Naz_99-1}
Maslov V.~P.   Nazaikinskii V. ~E. On the rate of convergence to the Bose-Einstein  distribution.
Math. Notes 2016;  \textbf{99} (1):95--109.

 \bibitem{MTN_98-1}
Maslov V.~P.  Case of less than two degrees of freedom, negative pressure, and the Fermi--Dirac distribution  for a hard liquid,''
Math. Notes 2015;  \textbf{98} (1):138--157.

\bibitem{MTN_102-2}
V.~P.~Maslov, `
`New Insight into the Partition Theory of Integers\protect\\
Related to Problems of Thermodynamics and Mesoscopic Physics,''
Math. Notes   \textbf{102}  (2)  234-- 251 (2017).

\bibitem{Ein_Born}
\textit{Albert~Einstein---Hedwig and Max~Born}, \textit{Briefwechsel} 1916--1955
(Nympherburger Verlagshandlung, M\"{u}nchen, 1969).

\bibitem{Litii}
 J.~ D.~Cockcroft and E.~T.~Walton, "Experiments with high velocity positive ions.
 II.
 The disintegration of elements by high velocity protons,"
 Proc.
 Royal Soc.
 A \textbf{137}, 229--242 (1932).

\bibitem{Bohr_Kal}
 N.~Bohr and F.~Kalckar ``On the transformation of atomic nuclei due
to collisions with material particles,'' Uspekhi Fiz.
 Nauk {\bf20} (3),
317--340 (1938).

\bibitem{Bohr_Wheeler}
 N.~ Bohr and J.
 A.
 Wheeler, ``The mechanism of nuclear fission,'' Phys.
 Rev. \textbf{56}, 426n450 (1939).


\bibitem{Lif_Pit-2}
 E.~M.~Lifshits and L.~P.~Pitayevskii, \emph{Theoretical Physics},
 Vol.~X: \textit{Physical Kinetics} (Fizmatlit, Moscow, 2007)
[in Russian].

\bibitem{Lif_Pit-1}
 E.~M.~Lifshits and L.~P.~Pitaevskii,
\emph{ Statistical Physics},   Vol.~X: Part~2: \textit{Theory of Condensed State}
(Nauka, Moscow, 1978; Pergamon, Oxford, 1980).

\bibitem{Auluck}
 F.~C.~Auluck and D.~S.~Kothari,
``Statistical mechanics and the partitions of numbers,''
 Math.
 Proc.
 Cambridge Philos.
 Soc. \textbf{42}, 272--277 (1946).

\bibitem{Agarwala}
 B.~K.~Agarwala and F.~C.~Auluck, ``Statistical mechanics and the partitions into non-integral powers of integers,''
 Math.
 Proc.
 Cambridge Philos.
 Soc. \textbf{47} (1), 207--216 (1951).

\bibitem{Rovenchak}
 A.~Rovenchak, \emph{Statistical Mechanics Approach in the Counting of Integer Partitions},
\texttt{\tt \texttt{arXiv:} 1603.01049v1 [math-ph] 3~Mar 2016}.

\bibitem{Frenkel}
 B.~Ya.~Frenkel,
\textit{Yakov Il'ich Frenkel}
(Nauka, Moscow--Leningrad, 1966)
[in Russian].

\bibitem{Abram}
\textit{Handbook of Mathematical Functions with Formulas, Graphs, and
Mathematical Tables},
 Ed. by M.~Abramowitz and I.~Stegun
(Dover Publications, New York, 1972;
 Nauka, Moscow,1979), pp. 825--826.

 \bibitem{RJMP_24-3}
V.~P.~Maslov, ``A Model of Classical Thermodynamics Based on the Partition Theory of Integers,
 Earth Garvitation, and Quasiclassical Assymptotics  I,''
Russian J.
Math.
Phys. \textbf{24} (3), 354--372   (2017).

\bibitem{Econ_arxiv}
 V.~P.~Maslov,
\textit{Threshold Levels in Economics},
\texttt{arXiv:0903.4783v2 [q-fin.ST], 3~Apr 2009}.


\bibitem{Teor_Vozm}
Maslov  V.~ P.
Perturbation Theory and Asymptotical Methods.
Moscow: Izd.
 Moskov.
 Univ.;1965;
Paris:  Dunod;1972
[in Russian and French].


\bibitem{Quantum_Evol}
Bayfield J.~E. Quantum Evolution.
New York--Toronto: John Wiley and Sons; 1999.

\bibitem{Theor_Optics}
R\"{o}mer H. Theoretical Optics.  Wienheim: Wiley-VCH; 2005.


\bibitem{Mas_Fed_Semi-Clas}
Maslov V.~P.,  Fedoriuk M.~V. Semi-Classical Approximation  in  Quantum  Mechnics.
Dordrecht-Boston-London:  D. Reidel Publishung Company;  1981.

\bibitem{MTN_99-2}
Maslov V.~P. On the Van-der-Waals forces.
Math. Notes 2016; 99 (2):284--289.


\bibitem{MTN_97-3}
Maslov V.~P. Gas--amorphous solid and liquid--amorphous solid
 phase transitions.
Introduction of negative mass and pressure from the mathematical viewpoint.
Math. Notes 2015; 97 (3):423--430.


\bibitem{Temperly}
Temperly H.~N. // Proc. Phys. Soc. (L.)  1947; \textbf{59}:199--205.

\bibitem{Burshtein}
Burshtein  A.~I.
Molecular Physics.
 Novosibirsk: Nauka;
1986
[in Russian].

\bibitem{Percival}
Percival  I. ~C. Semiclassical theory of bound states.  Advances in Chemical Physics, \textbf{36},
edited by Ilya Prigogine and Stuart A.  Rice.
New-York-London-Sydney-Toronto: John Wiley and Sons; 1977: 1--62.


\bibitem{RJMP_16-1}
Maslov  V.~ P.
Theory of chaos and its
application to the crisis of debts and the origin of the
inflation.
 Russian J.
 Math.
 Phys. 2009;
\textbf{16}
(1):103--120.


\bibitem{Anselm}
Anselm A. I. Foundations of Statistical Physics
and
Thermodynamics.
Moscow: Nauka; 1973
[in Russian].

\bibitem{Eyring}
Eyring H. J.,  Lin S.H., Lin S. M. Basic chemical kinetics.
New York: John Wiley and Sons Inc; 1980.

 \bibitem{Ram}
G.~H.~Hardy and S.~Ramanujan,
``Asymptotic formulae in combinatorial analysis,''
Proc. London Math. Soc.~(2) \textbf{17}, 75--115 (1917).

\bibitem{MTN_102-4}
V.~P.~Maslov,
``The Bohr--Kalckar correspondence principle
and a new construction of partitions in number theory,''
Math. Notes   \textbf{102}  (4),  533-- 540 (2017).


\bibitem{MTN_102-6}
V.~P.~Maslov,
``Two first principles of Earth surface Thermodynamics.
Mesoscopy, energy accumulation, and the
branch point in boson--fermion transition,''
Math. Notes   \textbf{102}  (6),   (2017).

\bibitem{RJMP_24-2}
V.~P.~Maslov, ``Topological phase transitions in the theory of partitions of integers,''
Russian J. Math. Phys. \textbf{24} (2), 249--260 (2017).


\bibitem{Postnikov}
 A.~G.~Postnikov,
\textit{Introduction to Analytic Number Theory}
(Nauka, Moscow, 1971).

\bibitem{Masl_Naz_100-3}
V.~P.~Maslov and V. ~E. ~Nazaikinskii,
``Conjugate variables in analytic number theory. Phase space and Lagrangian manifolds,''
Math. Notes  \textbf{100} (3), 421--428 (2016).

\bibitem{Masl_Dobr_Naz_2016}
V.~P.~Maslov, S.~Yu.~Dobrokhotov, and V.~E.~Nazaikinskii,
``Volume and entropy in abstract analytic number theory and thermodynamics,''
Math. Notes \textbf{100} (6), 828--834 (2016).

\bibitem{Masl_lovushka}
V.~P.~Maslov,
``Asymptotics of eigenfunctions of equations with boundary conditions on equidistant curves
and the distance between electromagnetic waves in the waveguide,''
Dokl. Akad. Nauk SSSR \textbf{123} (4), 631--633  (1958).

\bibitem{MTN_102-4_sh_com}
V.~P.~Maslov,
``Bounds of the repeated limit for the Bose--Einstein distribution and the construction
of partition theory of integers,''
Math. Notes \textbf{102} (4),  583--586 (2017).

\bibitem{9}
I.~A.~Kvasnikov, {\it Thermodynamics and Statistical Physics:
Theory of Equilibrium Systems} (URSS, Moscow, 2002), Vol.~2 [in Russian].

\bibitem{MTN_103-6}
V.~P.~Maslov,
``Statistical transition of the Bose gas to the Fermi gas,''
Math. Notes \textbf{103} (6), 3--9  (2018).

 \bibitem{Nestandart-1}
Robinson, A. Non-standard analysis.
North-Holland Publishing Co., Amsterdam 1966.

\bibitem{Nestandart-2}
Kanovei V.V., Reeken M.
 Nonstandard Analysis, Axiomatically.
  Springer, 2004.

\bibitem{Shepin-1}
E. V. Shchepin, ``Summation of Unordered Arrays,''
Funct. Anal. Appl. \textbf{52} (1), 35--44  (2018).

\bibitem{Shepin-2}
E. V. Shchepin, ``The Leibniz differential and the Perron--Stieltjes integral,''
J. Math. Sci. \textbf{233} (1), 157--171  (2018).

\bibitem{RJMP_25-2}
V.~P.~Maslov,
``Analytical number theory and the energy of
transition of the Bose gas to the Fermi gas.
Critical lines as boundaries of the
noninteracting gas (an analog of the Bose gas)
in classical thermodynamics,''
Russian J. Math. Phys.,
\textbf{25}
(2),
220--232
(2018).


\bibitem{14}
V.~P.~Maslov, ``A model of classical thermodynamics
and mesoscopic physics
based on the notion of hidden parameter, Earth gravitation, and
quasiclassical asymptotics. II,''
Russian J.
Math.
Phys. \textbf{24} (4), 494--504   (2017).

\bibitem{15}
V.~P.~Maslov,
``On the hidden parameter in  quantum and  classical  mechanics,''
Math. Notes \textbf{102} (6), 890--893   (2017).

\bibitem{16}
R.~Feynman,
\emph{Feynman Lectures in Physics}
(Librokom, 2015).

\bibitem{Bell}
J.~S.~Bell, ``On the Einstein Podolsky Rosen paradox,''
Physics \textbf{1} (3),  198--200 (1964).

\end{thebibliography}
\end{document}